\documentclass[aps,superscriptaddress,amsmath,amssymb,twocolumn,amsfonts]{revtex4-1}
\usepackage{amsmath,array}

\usepackage[varg]{txfonts}
\usepackage{graphicx}

\usepackage{hyperref}
\usepackage{cleveref}
\usepackage{color}
\usepackage{float}

\usepackage[normalem]{ulem}
\usepackage[resetlabels]{multibib}

\usepackage{xr}
\newcommand*{\myexternaldocument}[1]{%
	\externaldocument{#1}%
	\addFileDependency{#1.aux}%
}

\begin{document}
\title{Inducing and controlling superconductivity in Hubbard honeycomb
model using an electromagnetic drive}
\author{Umesh Kumar}
\affiliation{Theoretical Division, T-4, Los Alamos National Laboratory, Los Alamos, New Mexico 87545, USA}
\author{Shi-Zeng Lin}
\affiliation{Theoretical Division, T-4 and CNLS, Los Alamos National Laboratory, Los Alamos, New Mexico 87545, USA}
\date{\today}
\begin{abstract}
The recent successful experimental observation of quantum anomalous Hall effect in graphene under laser irradiation demonstrates the feasibility of controlling single particle band structure by lasers. Here we study superconductivity in a Hubbard honeycomb model in the presence of an electromagnetic drive. We start with Hubbard honeycomb model in the presence of an electromagnetic field drive, both circularly and linearly polarized light and map it onto a Floquet $t$-$J$ model. We explore conditions on the drive under which one can induce superconductivity (SC) in the system. We study the Floquet $t$-$J$ model within the mean-field theory in the singlet pairing channel and explore superconductivity for small doping in the system using the Bogoliubov-de Gennes approach. We uncover several superconducting phases, which break lattice or time reversal symmetries in addition to the standard $U(1)$ symmetry. We show that the unconventional chiral SC order parameter ($d \pm id$) can be driven to a nematic SC order parameter ($s+d$) in the presence of a circularly polarized light. The $d+id$ SC order parameter breaks time reversal symmetry and is topologically nontrivial, and supports chiral edge modes. We further show that the three-fold nematic degeneracy can be lifted using linearly polarized light. Our work, therefore, provides a generic framework for inducing and controlling SC in the Hubbard honeycomb model, with possible application to graphene and other two-dimensional materials.
\end{abstract}

\maketitle

\section{Introduction}
Graphene has been one of the most studied materials with hexagonal geometry due to its number of intriguing properties~\cite{Novoselov666, Geim2007, RevModPhys.81.109}. Recently, superconductivity was observed in a twisted bilayer at the magic angle~\cite{Cao2018}, after the theoretical prediction of the existence of flat bands for Moir\'e lattice of the twisted bilayer graphene at the magic angle~\cite{Bistritzer12233}. Superconductivity has also been observed in intercalated graphite such as ${\mathrm{CaC}}_{6}$~\cite{PhysRevLett.95.087003}, and C$_6$Yb~\cite{Weller2005}. But, in pristine graphene, superconductivity is still missing in spite of a number of theoretical studies predicting the existence of superconductivity~\cite{PhysRevB.75.134512,PhysRevLett.100.146404, Profeta2012,Nandkishore2012}.  Ref.~\cite{Nandkishore2012} predicted chiral superconductivity with nontrivial topology at van Hove singularity, which was an experimental challenge. More recently, doping graphene at and beyond van Hove singularity the have been achieved~\cite{PhysRevLett.125.176403} and can possibly open new venues for exotic states.  Many of these studies start with the assumption that honeycomb lattice has is a spin-$1/2$ antiferromagnet at half-filling, which has, in fact been observed in another honeycomb material, In$_3$Cu$_2$VO$_9$~\cite{PhysRevB.78.024420,PhysRevB.85.085102}. More recently, valence bond fluctuations were reported in another spin-1/2 honeycomb compound, YbBr$_3$~\cite{CWessler2020}. On the other hand, tunable honeycomb lattice in the cold atom system has been realized~\cite{Tarruell2012}. 

In recent years, there has also been an enormous interest in understanding the periodically driven systems. 
A recent study on graphene~\cite{McIver2020} reported the light-induced anomalous quantum Hall effect by tuning its band structure using light ~\cite{PhysRevB.79.081406}. Controlling the magnetic interaction in materials using an electromagnetic drive has been intensively studied
\cite{PhysRevLett.76.4250, Mentink2015, PhysRevLett.108.057202, RevModPhys.82.2731,Li2013, doi:10.1146/annurev-conmatphys-031218-013423}. The earlier work on photo-manipulation has largely focused on ferromagnets. But in recent years, study of antiferromagnets driven by light have also gained significant interest~\cite{Nemec2018}.

A recent study~\cite{PhysRevB.96.195110} on hexagonal lattice proposed the realization of light-induced time-reversal symmetry broken (TRSB) topological superconductor. In this study, the effect of electromagnetic drive was studied using a Pierls substitution and did not delve into how the frequency restricts the realization of the relevant Floquet $t$-$J$ model. Also, in one of the most studied superconductors, $i.e$ cuprate, there has been significant interest in enhancing superconductivity in the presence of a drive~\cite{Fausti189, Mankowsky2014} resonant with phonon mode, and the mechanism it is usually understood as phonon-assisted. More recently~\cite{PhysRevX.10.011053}, it has been shown that the enhancement in the superconductivity can be observed even when one is away from the resonance with the phonon. The mechanism, in this case, has been attributed to the effect of light on the superexchange coupling.  

The observation of band tuning in graphene~\cite{McIver2020} and the possibility of light-induced superconductivity mediated by charge transfer~\cite{PhysRevX.10.011053} opens new avenues for exploring light-induced superconductivity. Additionally, it has been widely reported in the literature that strong correlations can induce superconductivity in honeycomb lattice at low doping~\cite{Black_Schaffer_2014, PhysRevB.87.094521} and van Hove singularity~\cite{Nandkishore2012}. 
In our work, we explore the conditions for realizing superconductivity in honeycomb lattice mediated by strong correlations that can be controlled using light. We explore the effect of both the circularly and linearly polarized light on the electronic states. We show that in the frequency limit ($t\ll \omega \ll U$. Here $t$ is the hopping strength and $U$ is the onsite Coulomb interaction.), the superconductivity can be induced and enhanced. The honeycomb lattice is shown to host both the time reversal symmetry breaking (TRSB) ($d + i d$) and the nematic ($s+d$) superconductor in the singlet pairing channel. The nematic superconducting order parameter is three-fold degenerate. We further show that the three-fold degeneracy can be lifted using linearly polarized light. 

The paper is organized as follows: Sec~\ref{sec:FloquettJ} presents the Floquet $t$-$J$ model derived from Hubbard honeycomb model in the presence of circularly and linearly polarized light. Sec.~\ref{Sec:NonIntHC} discusses the density of states for the lattice in the presence of drive. 
Sec.~\ref{Sec:MFTtJ} discusses the mean-field treatment of Floquet $t$-$J$ model, which reveals several superconducting phases in the presence of EM drive. Sec.~\ref{sec:DiscConclusion} presents a discussion on the physical challenges in realizing the these Floquet systems and summarizes the various findings of our work.  

\section{Floquet Hamiltonian for Strongly correlated systems}\label{sec:FloquettJ}
Strongly correlated materials can be modeled using Hubbard model.
Time-dependence of the Hamiltonian in the presenc of a drive can be is taken into account by time-periodic Peierls phase~\cite{PhysRevB.96.195110, PhysRevLett.120.246402,Eckardt_2015}.  We start with a time-dependent Hamiltonian for a Hubbard honeycomb lattice given by 

\begin{equation}
\begin{split}
H(t) & =  \sum_{\langle ij \rangle,\sigma} t_{ij} e^{i\delta F_{ij}(t)}  c_{i\sigma}^\dagger c_{j\sigma}+  \text{h.c.}  + U \sum_{i} n_{i\uparrow} n_{i\downarrow}.
\end{split}
\end{equation}

Here, $t_{ij}$ is the hopping between nearest neighbor sites-$i, j$, and $U$ is the onsite electron repulsion. $\delta F_{ij}(t) =\textbf{F}(t) \cdot (\textbf{r}_i-\textbf{r}_j) $, where $\textbf{F}(t)$ is the vector potential of the light. The above time-dependent Hubbard Hamiltonian can be mapped onto a Floquet $t$-$J$ model dependent using Schrieffer-Wolff Transformation (SWT)~\cite{PhysRevLett.116.125301, PhysRevB.37.9753}. The details for the SWT are discussed in the supplemental material~\cite{Supplemental}. 

We investigate the Floquet $t$-$J$ model for two types of drive, $\textbf{ F}(t)$:  a) circularly polarized light (CPL)  and b) linearly polarized light (LPL) in the supplemental material~\cite{Supplemental}. The time-dependent Hamiltonian in the high frequency limit ($t/\omega \ll 1$), and $t\ll U$  can be mapped onto a set of time-independent Floquet $t$-$J$ Hamiltonians depending on the drive frequency ($\omega$)~\cite{Eckardt_2015, PhysRevLett.116.125301}. 
Here, we present the Floquet Hamiltonian for both circularly and linearly polarized light in the discussion below.\\
\begin{figure*}
	\centering
	\includegraphics[width=0.82\linewidth]{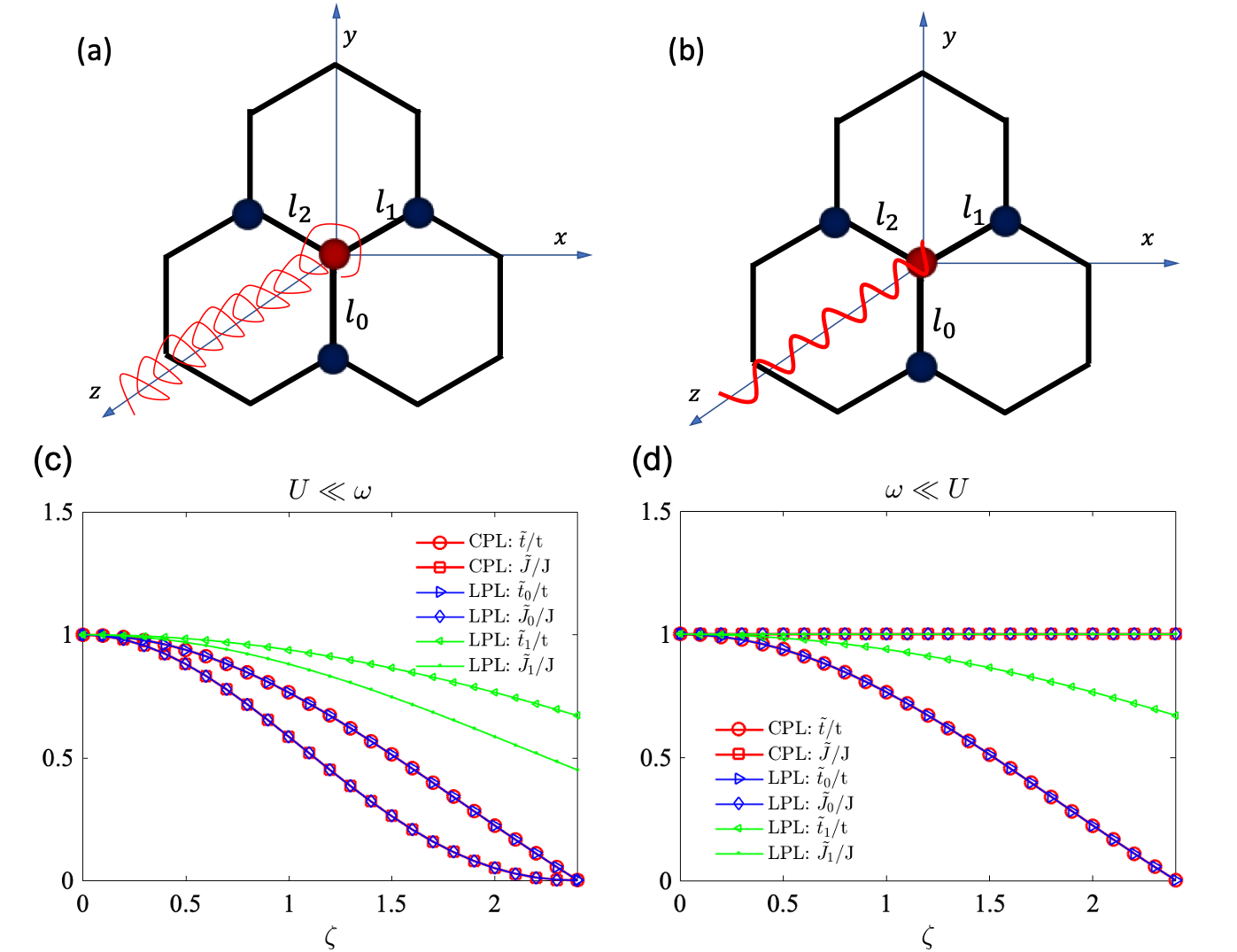}
	\caption{Schematics for the honeycomb lattice in the presence of (a) Circularly polarized light (CPL), and (b) Linearly polarized light (LPL) along $y$-direction. (c) and (d) show the rescaling of the Floquet hopping ($\tilde{t}$) and superexchange ($\tilde{J}$) with the drive parameter $\zeta (=A/\omega) $ in the limit $U\ll \omega$ (c) and  $\omega\ll U$ (d).}
	\label{fig:honeycombrstj}
\end{figure*}

{\sl Circularly Polarized Light:---} We consider a vector potential given by $\textbf{F}(t) =  \zeta[\sin(\omega t) \hat{e}_x - \cos(\omega t) \hat{e}_y]$, where $\zeta =A/\omega$ for the circularly polarized light (CPL) and $\hat{e}_x$ ($\hat{e}_y$) is a unit vector in the $x$ ($y$) direction.
In this case, the $D_\text{6h}$ point group symmetry of the Hamiltonian is preserved. This is because the CPL affects both the hopping and interaction isotropically along the three bonds in the hexagonal lattice.
We have two sets of Floquet Hamiltonian depending on the conditions on the frequency. 

In the limit $t\ll U \ll \omega$, the Floquet Hamiltonian is given by 
\begin{equation} 
\begin{split}
H_F  & \approx  t\mathcal{J}_0(\zeta ) \sum_{\langle ij\rangle, \sigma} \tilde{c}_{i,\sigma}^\dagger \tilde{c}_{j,\sigma} + \text{h.c.}\\
& + J \mathcal{J}_0^2(\zeta ) \sum_{\langle ij\rangle,\sigma}(\textbf{S}_i \cdot \textbf{S}_j - \sum_{\sigma'} \frac{1}{4}\tilde{n}_{i\sigma} \tilde{n}_{j\sigma'} ).
\end{split}
\end{equation} 
Here, $\tilde{c}_{j\sigma} = (1-n_{j\bar{\sigma}})c_{j\sigma} $, $\tilde{n}_{j\sigma} = (1-n_{j\bar{\sigma}})n_{j\sigma} $  and  $J=\frac{4t^2}{U}$ is the {\sl superexchange} interaction. $\mathcal{J}_0(\zeta)$ is a Bessel function of first kind with $\zeta = A/\omega$.  In this limit of $\omega$, both the hopping and {\sl superexchange}  are rescaled in the Floquet Hamiltonian as shown in the panel (b) of Fig.~\ref{fig:honeycombrstj}. 

On the other hand in the limit $t\ll \omega\ll U$, one can write the Floquet Hamiltonian as 
\begin{equation} \label{eq:CPLtwU}
\begin{split}
H_F  & \approx  t\mathcal{J}_0(\zeta ) \sum_{\langle ij\rangle,\sigma} \tilde{c}_{i,\sigma}^\dagger \tilde{c}_{j,\sigma} + \text{h.c.} \\
& + J\sum_{\langle ij\rangle, \sigma}  (\textbf{S}_i \cdot \textbf{S}_j - \sum_{\sigma'} \frac{1}{4}\tilde{n}_{i\sigma} \tilde{n}_{j\sigma'} ).
\end{split}
\end{equation} 
In this limit, only the hopping term is rescaled, whereas the {\sl superexchange} term remains unchanged. This limit can allows for flattening  the non-interacting band and whereas the {\sl superexchange} term interaction is constant, and hence an ideal condition for inducing and enhancing superconductivity.   

Additionally, we have chiral next-nearest neighbor (NNN) hopping terms of the order of $1/\omega$ that can be neglected owing to high-frequency approximation. Further, there are additional three site hopping terms that contribute at the same order as {\sl superexchange} in the high-frequency approximation of the Floquet theory~\cite{PhysRevB.37.9753, PhysRevLett.116.125301}. We further comment on the conditions under which these extra terms can be neglected in the supplemental~\cite{Supplemental}. \\

{\sl Linearly Polarized Light:---}  We consider a linearly polarized light (LPL) polarized along the $y$-direction for which the vector potential is given by   $\textbf{F}(t) =  \zeta\sin(\omega t) \hat{e}_y $, where $\zeta =A/\omega$. In this case, drive breaks the $C_\text{3}$-rotation symmetry along the three bonds in the Hamiltonian, and can lead to anisotropic hopping and interactions. As in the CPL case, here too, the Hamiltonian depends on the $U$ and $\omega$ strength. 

In the limit $t\ll U\ll \omega$ limit, the Floquet Hamiltonian is given by
\begin{equation} 
\begin{split}
H_F & \approx t \sum_{\langle ij\rangle, \sigma} \Big[ \mathcal{J}_0(\zeta ) \tilde{c}_{i_0\sigma}^\dagger \tilde{c}_{j\sigma} +\mathcal{J}_0(\tfrac{\zeta}{2}  ) \sum_{\ell=1, 2}  \tilde{c}_{i_\ell\sigma}^\dagger \tilde{c}_{j\sigma} + \text{h.c.}\Big]\\
& +J\sum_{\langle ij\rangle,\sigma} \Big[ \mathcal{J}_0^2(\zeta )  (\textbf{S}_{i_0} \cdot \textbf{S}_j - \sum_{\sigma'} \frac{1}{4}\tilde{n}_{i_0 \sigma} \tilde{n}_{j\sigma'} )\\
&+\mathcal{J}_0^2(\tfrac{\zeta}{2} )\sum_{\ell =1, 2} (\textbf{S}_{i_\ell} \cdot \textbf{S}_j - \sum_{\sigma'} \frac{1}{4}\tilde{n}_{i_2 \sigma} \tilde{n}_{j\sigma'} )  \Big].
\end{split}
\end{equation} 
In this case, both the hopping and superexchange are rescaled as is the case in CPL drive. In addition, the system develops anisotropy along the three bonds due to the explicit $C_3$ rotation symmetry breaking by the LPL.

On the other hand in the  limit $t\ll \omega\ll U$, the Hamiltonian is given by
\begin{equation} \label{eq:LPLtwU}
\begin{split}
H_F  & \approx t \sum_{\langle ij\rangle, \sigma} \Big[ \mathcal{J}_0(\zeta ) \tilde{c}_{i_0\sigma}^\dagger \tilde{c}_{j\sigma} +\sum_{\ell =1, 2} \mathcal{J}_0(\tfrac{\zeta}{2}  ) \tilde{c}_{i_\ell\sigma}^\dagger \tilde{c}_{j\sigma} + \text{h.c.}\Big]\\
& +J\sum_{\langle ij\rangle,\sigma}  (\textbf{S}_{i} \cdot \textbf{S}_j - \sum_{\sigma'} \frac{1}{4}\tilde{n}_{i \sigma} \tilde{n}_{j\sigma'} ).
\end{split}
\end{equation} 
In this case, the hopping is rescaled and develop anisotropy, whereas the superexchange is unaffected. It leads to flattening of the non-interacting band, whereas the superexchange term interaction is constant, and hence is useful for inducing and enhancing superconductivity

As in the CPL case, we do have NNN hopping terms, but the chiral hopping term is absent in the LPL drive as the Hamiltonian does not break time-reversal symmetry.  


\section{Non-interacting Honeycomb}\label{Sec:NonIntHC}
The effect of EM drive on non-interacting honeycomb lattice has been widely studied in literature, where the EM drive was found to have significant effects on electronic properties~\cite{PhysRevB.79.081406}. Here, we analyze the impact of drive on the honeycomb lattice without the superexchange term to prepare ourselves for understanding the more complicated Hamiltonian with strong correlations. We report the density of states (DOS) for non-interacting bands in the presence of both circularly and linearly polarized light. 

\begin{figure}
	\centering
	\includegraphics[width=\linewidth]{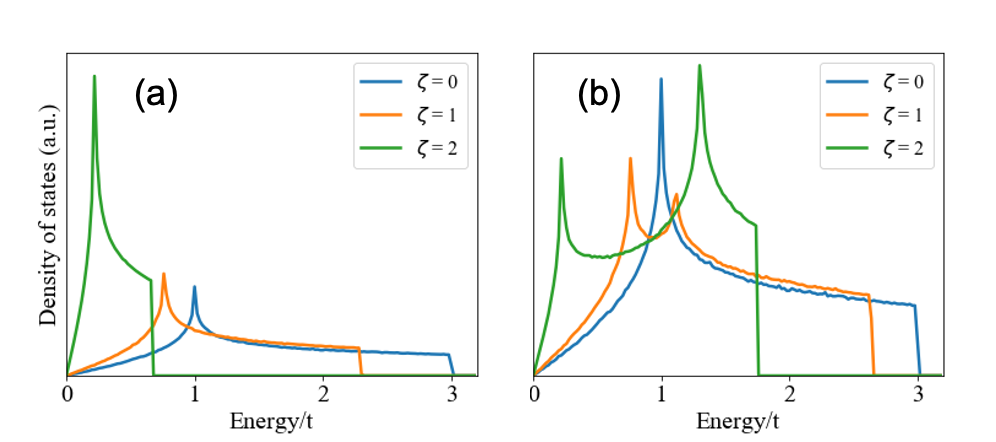}
	\caption{Density of states for non-interacting honeycomb lattice in the presence of; a) circularly polarized  light and b) linearly polarized along $y$-direction with the laser strength $\zeta$ indicated in each panel for $t\ll \omega$. The bandwidth gets smaller with increasing $\zeta$.}
	\label{fig:doscpl}
\end{figure}

{\sl Circularly polarized light:---} In the case of CPL, the hopping is renormalized isotropically and, therefore,  preserves the $C_\text{3}$-rotation symmetry of the bonds. 
Panel (a) in Fig.~\ref{fig:doscpl} shows the density of states for CPL drive. We can see that on driving the system ($\zeta$), only the bandwidth in the density of states shown in panel (a) of Fig.~\ref{fig:doscpl} gets smaller.

{\sl Linearly polarized light:---} Linearly polarized light can induce anisotropy in the hopping along with different bonds. It breaks $C_\text{3}$-rotation symmetry of the bonds and can generate effects on the honeycomb sites similar to the strain~\cite{Naumis_2017}. Panel (b) shows the DOS for the LPL. 

Additionally, CPL can breaks time-reversal symmetry in the Floquet $t$-$J$  model if the next nearest neighbor chiral hoppings are included. This term can lead to an anomalous quantum Hall effect (AQHE). The observation of light-induced AQHE in monolayer graphene was reported very recently~\cite{McIver2020}. This TRSB breaking term appears as a higher-order correction, and is only important when the leading order term proportional to the zeroth order Bessel function vanishes, $i.e.$ at $A/\omega = 2.4$~\cite{PhysRevB.93.144307}. For the sake of simplicity, we study the effects of drive in the limit of $\zeta<2$, which allows us to ignore the TRSB term.

\section{Mean-field treatement of the Floquet $t$-$J$ model}\label{Sec:MFTtJ}

In Sec.~\ref{sec:FloquettJ}, we have presented the Floquet Hamiltonian that can be engineered from a honeycomb Hubbard model. The derived $t$-$J$ Floquet model has an explicit many-body effect, which cannot be solved exactly. To simplify the many-body effects, one can use Gutzwiller projection to replace the strict double occupancy prohibition with a rescaling factor~\cite{Zhang_1988, Anderson_2004, PhysRevB.70.054504, Black_Schaffer_2014}. The Gutzwiller projected Hamiltonian could then be mapped to renormalized mean-field theory, using the mean-field for singlet pairing channel. In this renormalized mean-field theory, $t\rightarrow 2t\delta/(1+\delta)$ and $J\rightarrow 2J/(1+\delta)^2$ and superconducting order parameter, $\Delta \rightarrow 2\Delta\delta/(1+\delta)$, where $\delta ~(=1-n)$ is the doping away from half-filling. For the sake of simplicity, we will treat these rescaling factors as a constant. The Gutzwiller projected mean-field can be written as, 
\begin{equation}\label{eq:MFT}
\begin{split}
H_\text{MF} & =  -\sum_{\langle ij\rangle,\sigma} t_{ij} a_{i\sigma}^\dagger  b_{j\sigma} +\text{h.c.}  +\mu\sum_{i,\sigma} a_{i\sigma}^\dagger a_{i\sigma}+ b_{i\sigma}^\dagger b_{i\sigma} \\
& + \sum_{\langle ij\rangle,\sigma} (a_{i\uparrow}^\dagger b_{j\downarrow}^\dagger-a_{i\downarrow}^\dagger b_{j\uparrow}^\dagger) \Delta_{ij} + \Delta_{ij}^{*} (a_{i\downarrow} b_{j\uparrow} -a_{i\uparrow}b_{j\downarrow}).
\end{split}
\end{equation}
Here, $t_{ij}$ is the hopping between nearest neighbor sublattices, $A~(a_{i\sigma})$ and $B~(b_{i\sigma})$. Here we restrict to the  nearest neighbor singlet pairing superconducting order parameter (SCOP) $\Delta_{ij} = -J_{ij}(a_{i\downarrow} b_{j\uparrow}-a_{i\uparrow} b_{j\downarrow})/2$.


Equation ~\eqref{eq:MFT} can be written in the momentum basis as
\begin{equation}\label{eq:HamKspace}
\begin{split}
H_\text{MF}= &\sum_k
\begin{bmatrix}
a_{k \uparrow}^\dagger    b_{k \uparrow}^\dagger     & a_{-k \downarrow}   &  b_{-k \downarrow}    
\end{bmatrix}\times \\ &
\begin{pmatrix}
\mu       &   -t_{k}     &   0   &   \Delta_{k}       \\
-t_{k}^{*}      &   \mu       & \Delta_{-k}     &   0       \\
0  &  \Delta_{-k}^*   &   -\mu  &   t_{-k}^{*}  \\   
\Delta_{k}^*       &   0       & t_{-k}     &   -\mu       \\
\end{pmatrix}
\begin{bmatrix}
a_{k \uparrow}      \\
b_{k \uparrow}      \\
a_{-k \downarrow} ^\dagger  \\
b_{-k \downarrow} ^\dagger     \\
\end{bmatrix}.
\end{split}
\end{equation}
Here $t_\textbf{k} = \sum_{\alpha = 0, 1, 2} e^{i\textbf{k}\cdot \textbf{l}_\alpha }t_\alpha$ and $\Delta_\textbf{k} = \sum_{\alpha = 0, 1, 2} e^{i\textbf{k}\cdot \textbf{l}_\alpha }\Delta_\alpha$, where $\textbf{R}_j = \textbf{R}_i+\textbf{l}_\alpha$ with $l_0 = (0, -1), ~ l_1 = (\frac{\sqrt{3}}{2}, \frac{1}{2}), ~ l_2 = (-\frac{\sqrt{3}}{2}, \frac{1}{2})$. Also, one can write the gap as, $ \Delta_\alpha = -\frac{J_\alpha}{2}\sum_k \langle e^{i\textbf{k}\cdot\textbf{l}_\alpha} a_{-k\downarrow} b_{k\uparrow} - e^{-i\textbf{k}\cdot\textbf{l}_\alpha} a_{k\uparrow} b_{-k\downarrow} \rangle $. Equation~\eqref{eq:HamKspace} can be diagonalized to evaluate $H_{\text{MF}}|\psi_m\rangle = E_m |\psi_m \rangle$, the eigenstates ($|\psi_m \rangle$) of which are the Bogoliubov quasiparticles ~\cite{PhysRevB.98.195101}. The gap equation in this new basis is then given by 

\begin{equation}\label{Eq:Gap}
\begin{split}
 \Delta_\alpha & = \frac{J_\alpha}{4}\sum_{k,m} \langle e^{i\textbf{k}\cdot\textbf{l}_\alpha} u_{k\uparrow}^{a,m} (v_{-k\downarrow}^{b,m})^* \gamma_k^m (\gamma_k^m)^\dagger \\
 &+ e^{-i\textbf{k}\cdot\textbf{l}_\alpha} u_{k\uparrow}^{b,m} (v_{-k\downarrow}^{a,m})^* (\gamma_k^m)^\dagger \gamma_k^m \rangle.
 \end{split}
\end{equation}

 \begin{equation}\label{Eq:Doping}
 \begin{split}
  \bar{n} & = \sum_{k,m}  (|u_{k\uparrow}^{a,m}|^2 +  |u_{-k\downarrow}^{b,m}|^2 ) f(E_k^m) \\ &+(|v_{k\uparrow}^{a,m}|^2 +  |v_{-k\downarrow}^{b,m}|^2 ) f(-E_k^m).
 \end{split}
 \end{equation}
Here, $|\psi_k^m \rangle = \big(u_{k\uparrow}^{a,m},~v_{k \uparrow}^{b,m},~(u_{-k \downarrow}^{a,m})^*,~(v_{-k \downarrow}^{b,m})^*\big)^T$ and $E_k^m$ are
the $m^{\text{th}}$ eigenvector and eigenvalue of the Eq.~\eqref{eq:HamKspace}. $f(E_k^m) = \langle (\gamma_k^m)^\dagger \gamma_k^m  \rangle = \frac{1}{e^{\beta E_k^m}+1}$ is the Fermi distribution function for the Bogoliubov quasiparticles. We solve Eqs.~\eqref{Eq:Gap} and \eqref{Eq:Doping} self-consistently to evaluate the SC order parameter. 
 
\begin{figure}
	\centering
	\includegraphics[width=\linewidth]{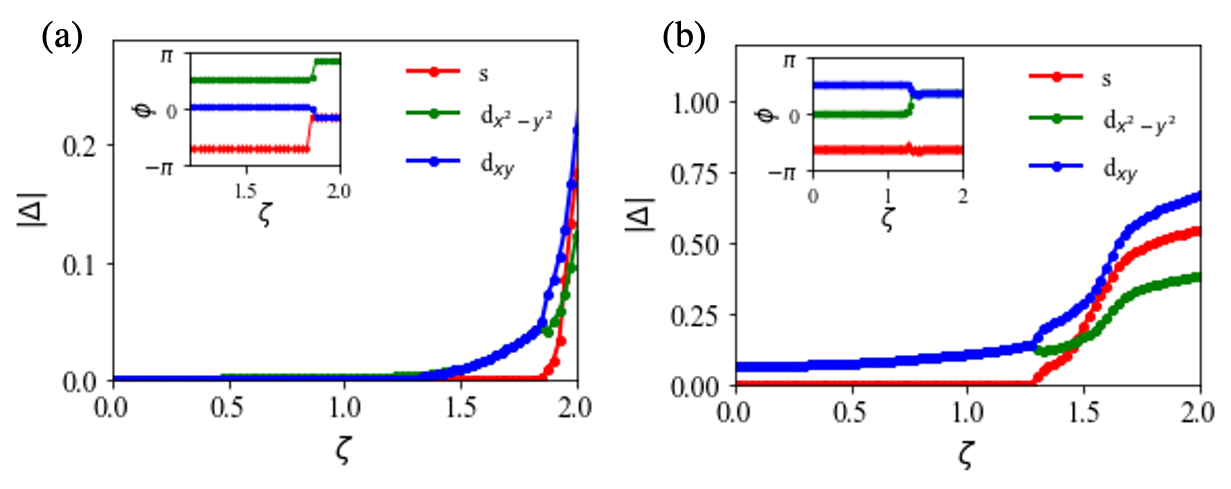}
	\caption{Superconducting order parameter (SCOP) as a function of CPL drive ($\zeta$). Panels (a) and (b) show the SCOP components for $J=1.0$,  $n=1.05$ and $J=2.0$,  $n=1.10$ respectively. Inset in each panels show the corresponding phase between the different component. Note that there is a global $U(1)$ phase ambiguity.}
	\label{fig:gapvszetat0}
\end{figure}

The basis functions; $\Psi_{d_{x^2-y^2}} = \tfrac{1}{\sqrt{6}}(2,-1,-1)$, $\Psi_{d_{xy}} = \tfrac{1}{\sqrt{2}}(0,1,-1)$ and $\Psi_s = \tfrac{1}{\sqrt{3}}(1,1,1)$ form a complete basis set for symmetric superconducting order parameter ($\Delta_\alpha$) for point group $D_\text{6h}$ and $D_\text{2h}$. Moreover, $d_{x^2-y^2}$ and $d_{xy}$ are degenerate for $D_\text{6h}$ point group, which allows for the existence of interesting chiral superconductivity $d_{x^2-y^2}+id_{xy}$. We evaluate the SCOP in these basis functions. All throughout the discussion, we set $t=1$ and all other parameters are defined in terms of $t$. 

\subsection{Superconducting order parameter dependence on the drive at $T=0$}
We start by discussing the superconducting order parameter (SCOP) dependence on the EM drive at $T=0$ for $t\ll \omega \ll U$. In this limit, the EM drive ($\zeta$) modifies the hopping parameter $t$. We plot SCOP for the circularly and linearly polarized light, respectively, for a set of $J$ and filling $n$ near the critical limit to induce different phases using the EM drive.

Figure~\ref{fig:gapvszetat0} plots the SCOP in the ($s, d$) basis in the presence of a circularly polarized light. The drive isotropically decreases the hopping ($t_\alpha = t\mathcal{J}_0(\zeta),~\forall~\alpha\in \{0,1,2\}$) along three bonds in the Floquet Hamiltonian as shown in  Eq.~\eqref{eq:CPLtwU}. The interaction parameter $J_\alpha~(= J,~\forall~\alpha\in \{0,1,2\})$ is unaffected. The Hamiltonian, therefore, preserves the $D_\text{6h}$ symmetry. Panel (a) plots the amplitude of the components of SCOP as a function of drive ($\zeta$) for $J= t~(U\approx 4t)$ and $n=1.05$, whereas panel (b) plots for  $J=2t~(U\approx 2t)$,  $n=1.1$. Inset in each panels show the phase of the different components along the different components of the SCOP. 

Panel (a) shows that one can drive the onset of superconductivity at $\zeta = 1.2$ using the EM drive. From the inset, it is clear that the initial SCOP is  $d_{x^2-y^2}+ i d_{xy}$, which spontaneously breaks the time reversal symmetry in addition to the $U(1)$ symmetry. The $d_{x^2-y^2}+ i d_{xy}$ is topological nontrivial and can be characterized by a Chern number $C=2$. The nontrivial topology implies the existence of spontaneous edge current. $d_{x^2-y^2}-i d_{xy}$ is another degenerate solution, guaranteed by time-reversal symmetry in the Hamiltonian.  There exists another phase transition at $\zeta =1.8$, associated with the onset of $s$-component. This transition is emphasized in panel (b). 

To further elucidate on the new transition, panel (b) plots SCOP for $J = 2t$ and $n=1.1$. In this parameter, at the onset of the drive, we have TRSB SCOP, signaled by $\pm \pi/2$ phase difference between different components of the SCOP. At  $\zeta = 1.2$, we see an onset of $s$-component. Further, in the inset, one can see that phase difference between the different components of the SCOP is either $0$ or $\pi$. The SCOP can be made real by choosing a global $U(1)$ phase, therefore the time reversal symmetry is restored above this critical $\zeta_c$. As will be discussed below, $C_3$ rotation symmetry is broken in this state, and therefore the superconductivity is nematic.\\

\begin{figure}
	\centering
	\includegraphics[width=\linewidth]{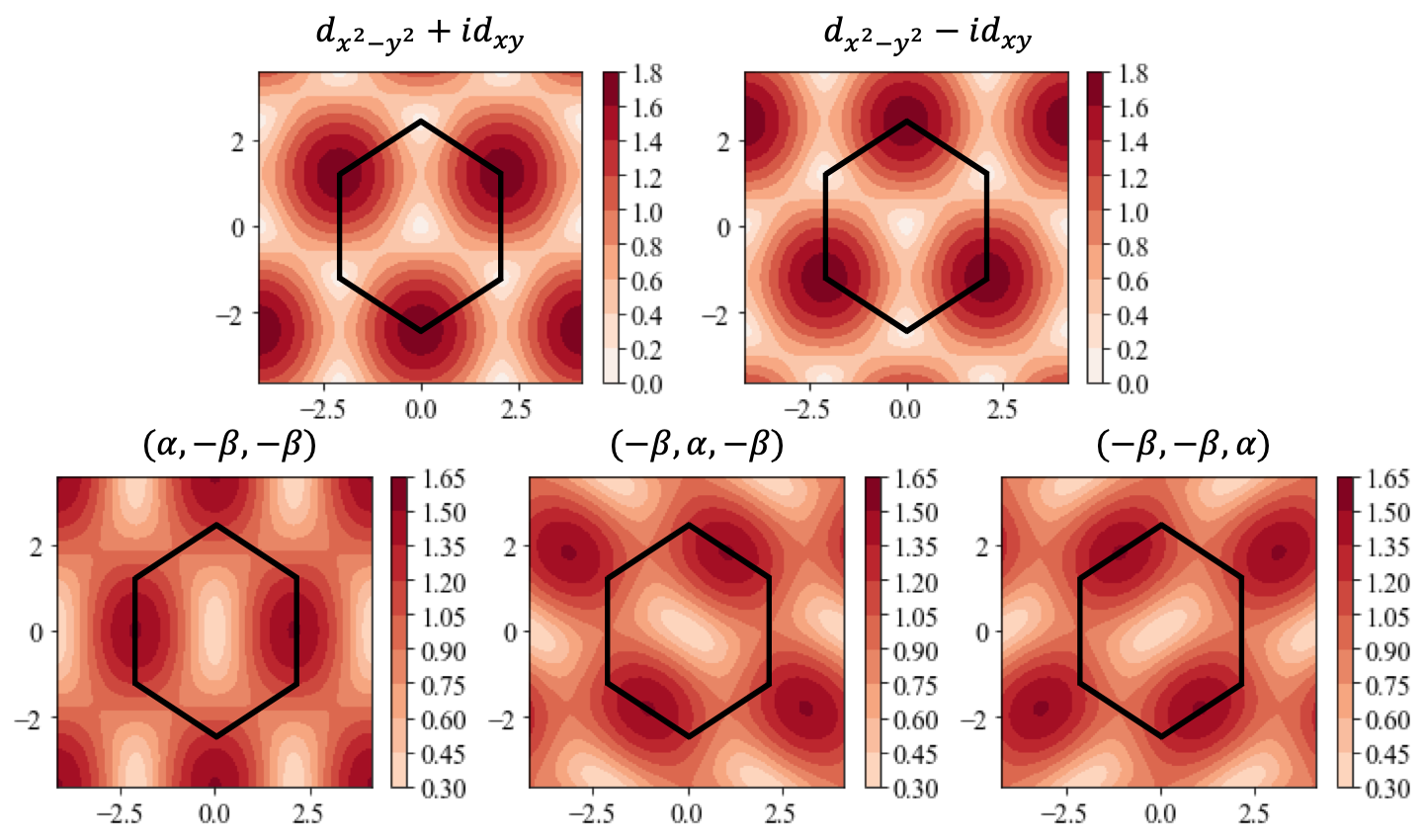}
	\caption{Top panels shows the superconducting order parameter (SCOP), $|\Delta (k)|$ for the time-reversal 
		symmetry-breaking; $d_{x^2-y^2}\pm i d _{xy}$. Bottom panels show SCOP for the three degenerate nematic states which breaks $C_{3}$ rotation symmetry along the three bonds. The Black hexagon denotes the first Brillouin Zone. 
	}
	\label{fig:gapkdependence}
\end{figure}

{\sl $\textbf{k}$-dependence of superconducting order parameter:---} We plot the momentum dependence of the SCOP, $\Delta(k) = \sum_k e^{i\textbf{k}\cdot R_\alpha} \Delta_\alpha$.  Fig.~\ref{fig:gapkdependence} shows the gap amplitude, $|\Delta(k)|$ distribution over the Brillouin Zone.  

SCOP, $d_{x^2-y^2}\pm i d_{xy}$, shown in panel (a) and (b) have point nodes and is not invariant for any global $U(1)$ phase over the time-reversal operation, $\mathcal{T } \Delta (\textbf{k})\rightarrow \Delta^* (-\textbf{k})$, and is indeed  TRSB SCOP. But this SCOP is invariant up to a global $U(1)$ phase over $C_\text{3}$ rotation and the SCOP components along the the three bonds $(\Delta_0, \Delta_1, \Delta_2)$ have the same amplitude. Note, while the SCOP has point node, the dispersion for the Bogoliubon quasiparticle is gapped.

\begin{figure*}
 	\centering
 	\includegraphics[width=\linewidth]{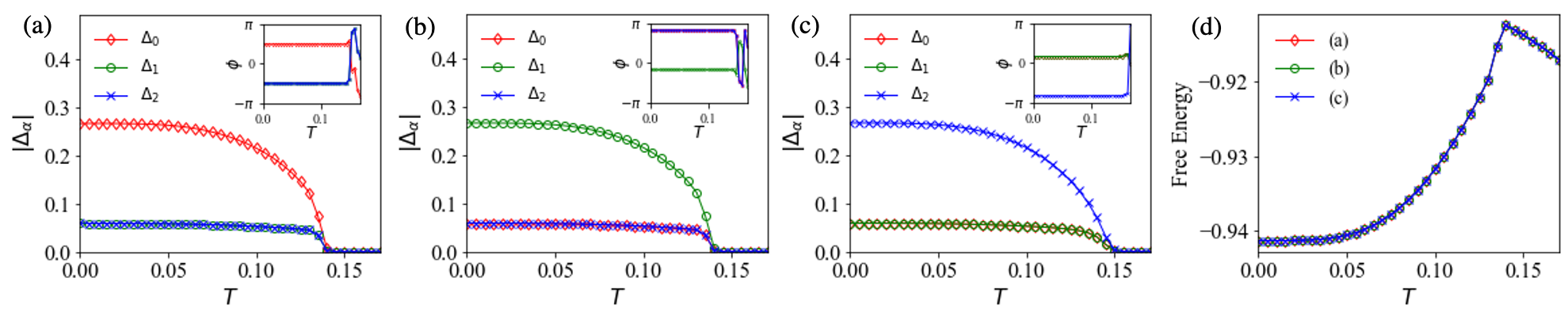}
 	\caption{Nematic SCOP degeneracy in the presence of CPL drive at $\zeta = 1.4$ for $J=2t$ and $n=1.1$. Panels (a), (b) and (c) plot the SCOP components along the three bonds $(\Delta_0, \Delta_1, \Delta_2)$ as a function of temperature for the three states. Panel (d) plots the free energy for the three states, showing state all the states are degenerate. 
 	}
 	\label{fig:nematicscvstemp}
 \end{figure*}

With the onset of $s$-wave component, the SCOP breaks the three-fold rotation symmetry $C_\text{3}$ and the SCOP component along one of the bond $(\Delta_0, \Delta_1, \Delta_2)$ becomes inequivalent, which indeed leads to nematic SCOP. There are three degenerate SCOPs that are related by $C_3$ rotation (This is further emphasized in Fig.~\ref{fig:nematicscvstemp}). In the nematic phase, the SCOP spontaneously breaks the $C_\text{3}$ point group symmetry and is reduced to $D_\text{2h}$. The SCOP
 has form $\Delta_\alpha \propto \exp(i\theta)(\alpha, -\beta, -\beta,),~\exp(i\theta)(-\beta, \alpha, -\beta),$ and $\exp(i\theta)(-\beta, -\beta, \alpha)~\forall~\alpha> \beta>0$.
Panel (c), (d), (e) plots the momentum dependence of the nematic SCOP for $\alpha=1$ and $\beta=1/2$. As can be seen in these panels, the nematic order parameter ($s+d$) is nodeless 
and preserves the time reversal symmetry, but breaks the  $C_\text{3}$ rotation symmetry.
Conversely, in the ($s,~d_{x^2-y^2},~d_{xy}$) basis, these nematic order parameters can be written as;  $\exp(i\theta)\big(\tfrac{\alpha-2\beta}{\sqrt{3}} s + \tfrac{2\alpha+2\beta}{\sqrt{6}} d_{x^2-y^2}\big)$, $\exp(i\theta)\big(\tfrac{\alpha-2\beta}{\sqrt{3}} s + \tfrac{-\beta-\alpha}{\sqrt{6}} d_{x^2-y^2} +  \tfrac{\alpha+\beta}{\sqrt{2}} d_{xy}  \big)$ and  $\exp(i\theta)\big(\tfrac{\alpha-2\beta}{\sqrt{3}} s + \tfrac{-\alpha-\beta}{\sqrt{6}} d_{x^2-y^2} +  \tfrac{-\beta-\alpha}{\sqrt{2}} d_{xy}  \big)$ respectively. Note, in the panel (b) of Fig.~\ref{fig:gapvszetat0}, we have plotted only one of the solution of the nematic SCOP, the other two solutions can be recovered by rotation. \\

\begin{figure}
	\centering
	\includegraphics[width=\linewidth]{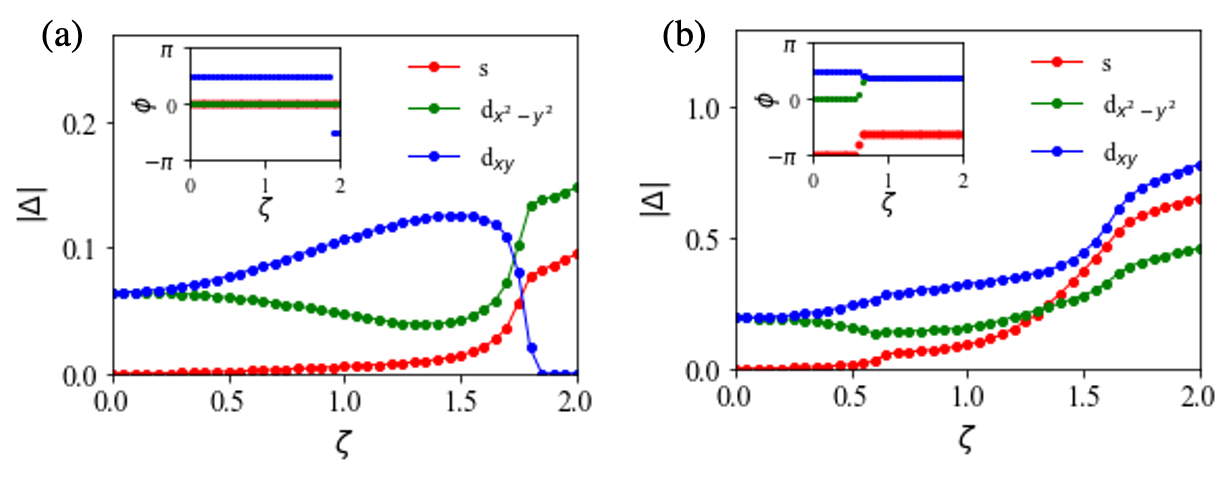}
	\caption{SCOP as a function of LPL drive ($\zeta$). Panels (a) and (b) show the SCOP components for $J=2.0$,  $n=1.10$ and $J=3.0$,  $n=1.10$ respectively. Inset in each panels shows the corresponding phase between the different component. Note that there is a global $U(1)$ phase ambiguity.}
	\label{fig:lplgapvszeta}
\end{figure}

Fig.~\ref{fig:lplgapvszeta} plots the SCOP in the presence of a linearly polarized light. For LPL polarized along $y$-direction, the parameters rescales as $i.e.$ $t_0 = t\mathcal{J}_0(\zeta), t_1 = t_2 = t\mathcal{J}_0(\zeta/2)$, whereas $J_\alpha = J,~\forall \alpha\in \{0,1,2\}$ as can be seen from Fig.~\ref{fig:honeycombrstj} (d). In $y$-polarized LPL, $D_\text{6h}$ symmetry of the Hamiltonian is reduced to $D_\text{2h}$. Panel (a) plots the amplitude of the components of SCOP as a function of drive ($\zeta$) for $J= 2t~(U\approx 2t)$ and $n=1.1$, whereas panel (b) plots for  $J=3t~(U\approx 8t/3)$,  $n=1.1$. Inset in each panels show the phase of the different components along the different components of the SCOP. Note, we have plotted for larger value of $J$ in this case as the onset of SCOP is slower in the case of LPL. Also, the SCOP in the case of LPL are distinct from the CPL case.

Panel (a) shows that in the case of $J= 2t~(U\approx 2t)$ and $n=1.1$, the SCOP is always complex for arbitrary choice of the $U(1)$ phase and is TRSB. We choose a global phase such that the real part is composed of $d_{x^2-y^2}$ and $s$, unlike in CPL where it consists of only $d_{x^2-y^2}$. From the inset, it is clear that the initial SCOP is indeed $\Psi_{sd}+ i d_{xy}$. $\Psi_{sd}-i d_{xy}$ is another degenerate solution, which is guaranteed by time-reversal symmetry in the Hamiltonian.
Here $\Psi_{sd}$ represents SCOP with the mixed $s$ and $d_{x^2-y^2}$.

 \begin{figure*}
 	\centering
 	\hspace*{-0.25in}
 	\includegraphics[width=1.0\linewidth]{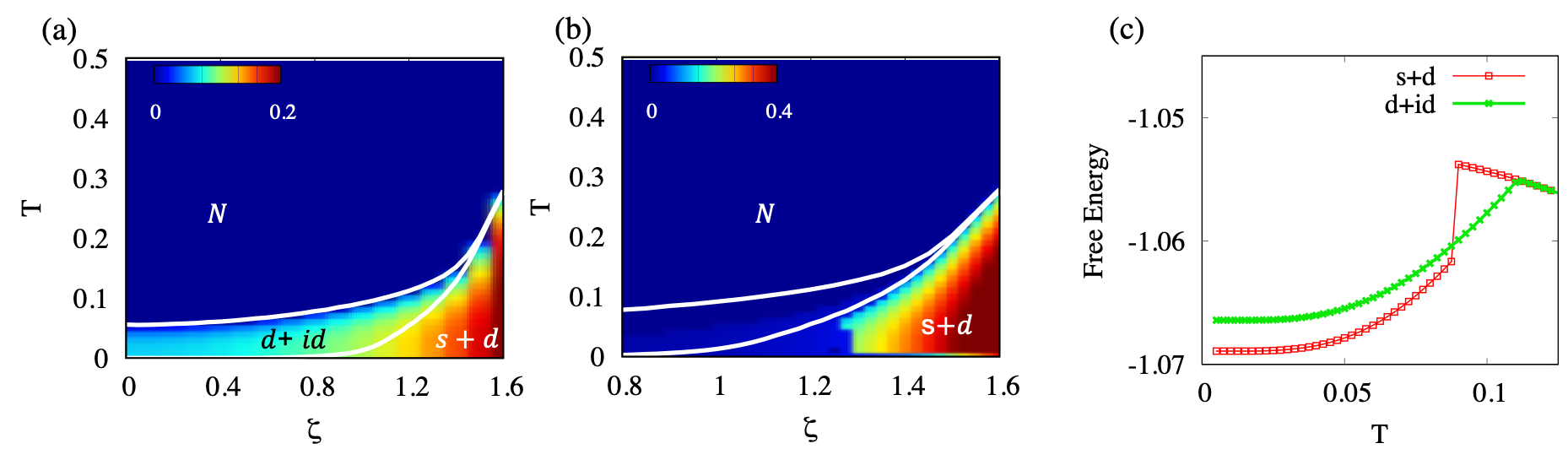}
 	\vspace*{-0.25in}
 	\caption{Phase diagrams for the SCOP as a function of temperature ($T$) and the CPL drive ($\zeta$) for $J=2.0,~n=1.1$ and $t=1$. Panels (a) and (b) show the SCOP evaluated using projected $d_{x^2-y^2}+id_{xy}$ and $s$ state respectively as initial guess. The top and bottom lines in each panel indicate the $T_c$ for vanishing $d$-wave and $s$-wave SCOP respectively evaluated using linearized gap equation. Panel (c) plots the free energy at $\zeta = 1.23$ and shows that nematic state with $(s+d)$-wave component is the true SCOP at lower temperature, and at intermediate temperature ($d+id$) dominates before the SCOP vanishes. }
 	\label{fig:gapvstvszeta}
 \end{figure*}

\begin{figure*}
\centering
\includegraphics[width=\linewidth]{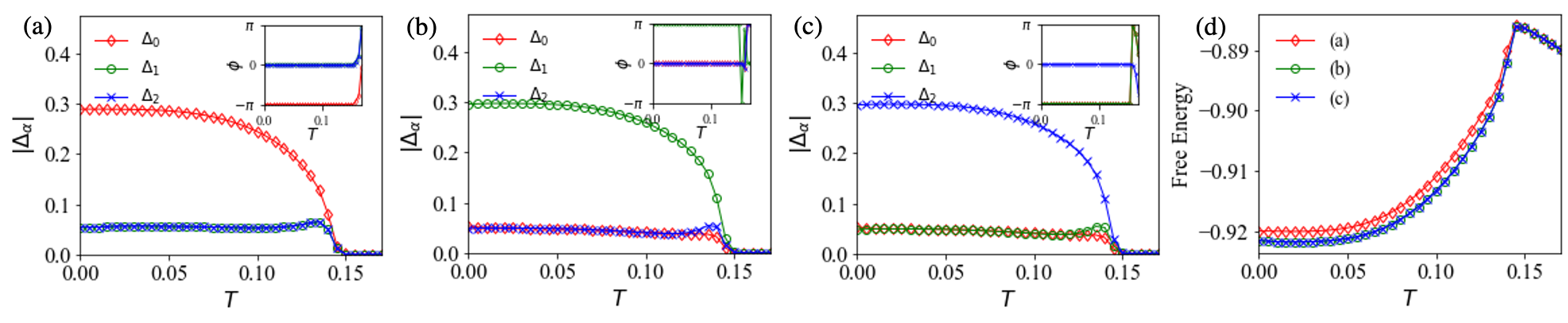}
 \caption{Nematic SCOP in the presence of LPL drive at $\zeta = 0.5$ for $J=2\tilde{t}$, $\tilde{t}=t\mathcal{J}_0(1.4)$ and $n=1.1$. Panels (a), (b) and (c) plot the SCOP components along the three bonds $(\Delta_0, \Delta_1, \Delta_2)$ as a function of temperature for the three states. Panel (d) plots the free energy for the three states, showing state in panel (a) becomes non-degenerate.
 }
\label{fig:LPLnematicscvstemp}
\end{figure*}

Panel (b) plot SCOP for larger $J=3t$ to enter into the nematic state observed in the case of CPL.  We notice that the SCOP becomes real at $\zeta = 0.5$, as was also observed in the CPL case. In this phase, the time-reversal symmetry of the SCOP is preserved. The three-fold degeneracy of the nematic state observed in the CPL case is reduced to two-fold degeneracy. This will be further evident from the discussion on Fig.~\ref{fig:LPLnematicscvstemp}. Again note, we have plotted only one of the nematic phase solutions in panel (b).

\subsection{Effect of temperature on the superconducting order parameter}
In this section, we discuss the effect of temperature on the superconducting order parameter for the CPL drive.  The vanishing SCOP reveals the critical temperature, $T_c$  for the phase transitions.  We further estimate the $T_c$ by linearizing the gap equations. We describe both the methods for evaluating the temperature dependence of SCOP. 

We evaluate the SCOP by solving the Eqs.~\eqref{Eq:Gap} and \eqref{Eq:Doping} self-consistently for finite temperature ($T$). We plot the phase diagram starting with a SCOP ansatz given by, $\Delta_\alpha  = |\Psi_{{d_{x^2-y^2}}+i{d_{xy}}}\rangle \langle \Psi_{{d_{x^2-y^2}}+i{d_{xy}}} | \Delta_\alpha \rangle $ to evaluate for TRSB state and $\Delta_\alpha  = |\Psi_{d_{s}} \rangle \langle \Psi_{d_{s}} | \Delta_\alpha \rangle $ for the nematic state.  
Since the TRSB and nematic order parameter competes, we also evaluate the free energy of the system~\cite{PhysRevB.58.9365} to get the true ground state. The free energy is evaluated using the relation,
 \begin{equation}
 \begin{split}
 \mathcal{F}&  =  \sum_{k,m} E_k^m f (E_k^m)+ T k_B \sum_{k,m} \Big(f (E_k^m) \ln(f (E_k^m)) \\
 & +\big(1- f (E_k^m)\big)\ln\big(1- f (E_k^m)\big)\Big)+2\sum_\alpha |\Delta_\alpha^2|/N.
 \end{split}
 \end{equation}
 Here $m$ is indices for the Bogoliubons evaluated in the Eq.~\ref{eq:HamKspace}.

We know that at the critical transition temperature, $T_c$, the SC gap vanishes, and therefore one can linearize the gap equation. This approximation allows us to directly calculate the $T_c$, which is discussed in Appendix~\ref{App:LinGap}. 
The linearized gap equation can be written in a  matrix form
\begin{equation}\label{eq:GapSelfConsistent}
 \begin{bmatrix}
 \Delta_{0}     \\
 \Delta_{1}\\
 \Delta_{2}   \\
 \end{bmatrix}
 =J \begin{pmatrix}
 A_1		& 	B_1		&  B_1	\\   
 B_1		& 	A_2		&  B_2	\\   
 B_1		& 	B_2		&  A_2	\\   
 \end{pmatrix}
 \begin{bmatrix}
 \Delta_{0}     \\
 \Delta_{1}\\
 \Delta_{2}   \\
 \end{bmatrix},
 \end{equation}
 where $A_1$, $A_2$, $B_1$ and  $B_2$ are components in the Eq.~\eqref{eq:lnrzdGapEq} when $J_\alpha = J,~\forall~\alpha\in\{0, 1, 2\}$, discussed in Appendix~\ref{App:LinGap}. In the case of CPL, $A_1=A_2= A$ and  $B_1=B_2= B$. 
 The solutions of the above equation is given by, a) $A+2B=\frac{1}{J} $ $s$-wave;  $(1,1,1)/\sqrt{3}$,   and b) $A-B=\frac{1}{J} $,  two-fold $d$-wave degenerate solutions: $(2,-1,-1)/\sqrt{6}$ and  $(0,1,-1)/\sqrt{2}$. Using these equations, we evaluate the $T_c$ for the $d$-wave and $s$-wave solution \\

\subsubsection{Phase diagram of SCOP dependence on temperature and CPL drive}\label{sec:PhaseDiagram}  
Figure~\ref{fig:gapvstvszeta} shows the phase diagram for the temperature dependence of the SCOP in the presence of CPL drive evaluated for $J=2t$ and $n=1.1$. We reveal distinct phase boundaries for the TRSB and nematic superconductivity. TRSB and nematic SCOP can compete at lower temperatures, and therefore, one has to compare the free energy in order to determine the true ground state.  

Panels (a) plots the  phase diagram with a mean-field gap ansatz given by, $\Delta_\alpha  = |\Psi_{d_{x^2-y^2} + id_{xy}}\rangle \langle \Psi_{d_{x^2-y^2}+i d_{xy}} | \Delta_\alpha \rangle $. We further evaluate that $T_c$ for the $d$-wave solution using the linearized gap equations and compare it with the results from the SCOP calculation with $(d+id)$ ansatz. We notice the TRSB SCOP vanishes at the $T_c$ (top white line) of the $d$-wave solution.

Panels (b) plots the  phase diagram with a mean-field gap ansatz given by, $\Delta_\alpha  = |\Psi_{d_{s}}\rangle \langle \Psi_{d_{s}} | \Delta_\alpha \rangle $. We further evaluate that $T_c$ for the $s$-wave solution using the linearized gap equations and compare it with the results from the SCOP calculation with $s$-wave ansatz. We notice the nematic SCOP vanishes at the $T_c$ (bottom white line) of the $s$-wave solution. We notice that there is a discrepancy in the critical drive strength for the onset of the $s$-wave state; $\zeta_c = 0.85$ from the $T_c$ calculation, whereas, $\zeta_c = 1.25$ from the SCOP calculation. This late onset of $s$-wave SCOP is due to numerical limitation, a smaller grid used for evaluating the SCOP.

From panel (a) and (b), it is clear that $(d+id)$ and $(s+d)$-wave solutions compete at a lower temperature after the onset of $s$-wave state. We, therefore, plot free energy for both the solution to find the real ground state. Panel (c) plots the free energy at $\zeta = 1.23$ to show that the $(s+d)$-wave solution is the real ground state at the lower temperature after the onset of $s$-wave. At a higher temperature, the $(d+id)$-wave solution has lower free energy after the $(s+d)$-wave solution vanishes.
 
Additionally, we notice that the $T_c$ for TRSB and nematic SCOP converge for a large $\zeta$. \\ 

{\sl Analysis of the three-fold degeneracy of nematic state:---} SCOP in the nematic state has a three-fold degeneracy. In Fig.~\ref{fig:nematicscvstemp}, we plot a cut at $\zeta =1.4$ from panel (b) of Fig.~\ref{fig:gapvstvszeta} to explore the temperature dependence of the components of SCOP along all the three bonds. We plot all the three nematic states shown in panels (a), (b) and (c) by biasing the initial SCOP along each of the bond. Inset plots the phases of the three components along the bonds.  
We notice that the SCOP has a higher amplitude along one of the bond in each state and are related by a $C_3$ rotation along the three bonds. 

We further plot the free energy for these three solutions in panel (d) and observe that these solutions' free energy is equal for all the temperature to confirm that these solutions are indeed degenerate. 

{\sl Breaking the three-fold degeneracy in the nematic state:---} We know that LPL reduces the $D_\text{6h}$ symmetry to $D_\text{2h}$ in the presence of $y$-polarized light. We use this property and show that the use of linearly polarized light can lift the three-fold degeneracy of the SCOP. For the sake of consistency with the above results, we start with $J=2t$, and a modified hopping $\tilde{t}=t\mathcal{J}_0(1.4)$, which reproduces the results in the case of CPL at $\zeta =1.4$. Additionally, we use a $y$-direction LPL with $\zeta=1/2$ to break the three-fold degeneracy,  as shown in Fig.~\ref{fig:LPLnematicscvstemp}. The three nematic states shown in panels (a), (b) and (c) are plotted by biasing the SCOP along each of the bonds. We notice that the SCOP in panel (a) has a slightly different amplitude from that in panel (b) and (c). 

To understand the true nature of the ground state, we plot the Free energy in panel (d). We notice that the free energy for polarization along the $l_0$ is higher than the other two directions. Therefore, the three-fold degeneracy of the SCOP is reduced to two-fold degeneracy in the presence of LPL.
 


\section{Discussion and Conclusions}\label{sec:DiscConclusion}
 In our work, we explored the conditions for enhancing superconductivity in strongly correlated honeycomb lattice in the presence of both circularly and linearly polarized light. An earlier work~\cite{PhysRevB.96.195110} on honeycomb lattice also explored the same system by treating EM drive through a Pierls-substitution. On the other hand, we have treated the effect of strong correlation systematically using SWT and find that the limit $t\ll \omega \ll U$ is useful in enhancing superconductivity. We want to point out that two conditions are important for realizing such enhancement; a) $t\ll \omega \ll U$, such that the higher-order corrections in the high-frequency approximations can be neglected, and b) thermal heating is minimal to realize Floquet Hamiltonian. 
To have a minimal effect on the higher-order correction, we study the system for $\zeta<2$. In driven systems, a number of studies have reported the effect of heating and shown that, indeed, it can be controlled. For example, it has been shown that the Floquet heating in the many-body systems in the high frequency $t, U<\omega$ is exponentially slow in frequency ~\cite{PhysRevLett.115.256803}. Further, it has been shown that Floquet prethermalized state can be realized in a resonantly driven Hubbard model~\cite{Herrmann_2017, PhysRevB.96.085104}.
Additionally, the recent experimental observation of off-resonance superconductivity enhancement in cuprate~\cite{PhysRevX.10.011053}, possibly mediated by charge fluctuations, provides further impetus to our work. 


In conclusion, we have shown that light can be a useful tool for controlling the bandwidth and strong correlations in a hexagonal lattice. We show that in the frequency limit, $t<\omega$ and  $\omega<U$, the EM field drive can induce and enhance superconductivity mediated through strong correlations. We show that light-induced superconductivity is exotic; a)  time-reversal symmetry breaking ($d+id$) with nontrivial topology and b) nematic ($s+d$) in these honeycomb antiferromagnets. We also show how the TRSB state can be driven into a nematic superconductor in the presence of both; circularly and linearly polarized light. 
Our study also explores the effect of temperature on the SCOP and presents a phase diagram for the temperature dependence of the SCOP in the presence of a CPL drive. On the discussion of the temperature of dependence, we also present all the three-fold degenerate nematic state and confirm their degeneracy using a calculation of the free energy.   
We further show that this three-fold degeneracy of the nematic state can be reduced to a two-fold in the presence of an LPL.
Our work, therefore, presents a detailed analysis of exploring the superconductivity mediated through a strong correlation in the presence of electromagnetic drive. Though our work explored the driven honeycomb lattice, but is not limited and is applicable to generic strongly correlated systems.

\section{Acknowledgements} The authors thank Ying Su for helpful discussion at the initial stage of the work. Computer resources for numerical calculations were supported by the Institutional Computing Program at LANL. This work was carried out under the auspices of the U.S. DOE NNSA under contract No. 89233218CNA000001 through the LDRD Program. S.-Z. L was also supported by the U.S. Department of Energy, Office of Science, Basic Energy Sciences, Materials Sciences and Engineering Division, Condensed Matter Theory Program.

\onecolumngrid

\begin{appendix}
\section{Evaluating $\textbf{T}_c$ by linearizing the Gap equation}\label{App:LinGap}
Here we present an analysis of the Hamiltonian to further investigate the nature of the superconductivity in the lattice. We map the bipartite lattice to a band basis. Following the work in Ref.~\cite{PhysRevB.75.134512, Black_Schaffer_2014},  one can transform the Hamiltonian in Eq.~\eqref{eq:HamKspace} to intraband and interband pairing, using the basis transformation:
 \begin{equation}
 \begin{pmatrix}
 a_{\textbf{k}\sigma}      \\
 b_{\textbf{k}\sigma}         \\
 \end{pmatrix}
 = \frac{1}{\sqrt{2}}\begin{pmatrix}
 c_{\textbf{k}\sigma} +d_{\textbf{k}\sigma}     \\
 e^{-i\phi_\textbf{k}}(c_{\textbf{k}\sigma}   -d_{\textbf{k}\sigma} )    \\
 \end{pmatrix}
 \end{equation}
 
 And the Hamiltonian is terms of this new basis given by 
 \begin{equation}\label{eq:HamIntraInter}
 \begin{split}
 H_\text{MF} = &  \sum_k
 \begin{bmatrix}
 c_{k \uparrow}^\dagger   & c_{-k \downarrow}   & d_{k \uparrow}^\dagger   &  d_{-k \downarrow}       
 \end{bmatrix} 
 \begin{pmatrix}
 \xi_1     &   \Delta_{i}     &   0   &   -\Delta_{I}       \\
 \Delta_{i}^\dagger       &   -\xi_1       & \Delta_{I}^\dagger     &   0       \\
 0  &  \Delta_{I}   &   \xi_2     &   -\Delta_{i}  \\   
 -\Delta_{I}^\dagger       &   0       &-\Delta_{i}^\dagger     &   -\xi_2     \\
 \end{pmatrix}
 \begin{bmatrix}
 c_{k \uparrow}     \\
 c_{-k \downarrow}^\dagger \\
 d_{k \uparrow}    \\
 d_{-k \downarrow}^\dagger     \\
 \end{bmatrix}
 \end{split}
 \end{equation}
 
 Here $\xi_1 = \mu- \epsilon_\textbf{k}$, $\xi_2 = \mu+\epsilon_\textbf{k}$ where $\epsilon_\textbf{k} = |\sum_{\alpha}t_\alpha e^{i\textbf{k}\cdot \textbf{R}_\alpha}|$ and $\phi_\textbf{k} = \text{arg}(\sum_{\alpha} e^{i\textbf{k}\cdot \textbf{R}_\alpha} t_\alpha)$. $\Delta_{\alpha} = J_\alpha\sum_{\textbf{k}} \cos(\textbf{k}\cdot\textbf{R}_\alpha-\phi_\textbf{k}) \big(c_{k \uparrow} c_{-k \downarrow} +d_{k \uparrow} d_{-k \downarrow}\big)+i\sin(\textbf{k}\cdot\textbf{R}_\alpha-\phi_\textbf{k}) \big(c_{k \uparrow} d_{-k \downarrow} - d_{k \uparrow} c_{-k \downarrow} \big) $. Further, we rewrite the gap as, intraband gap, $\Delta_{i}(\textbf{k}) = \text{Re}[\sum_{\alpha}e^{i(\textbf{k}\cdot\textbf{R}_\alpha-\phi_\textbf{k}) }\Delta_\alpha]$, and interband gap, $\Delta_{I}(\textbf{k}) = \text{Im}[\sum_{\alpha}e^{i(\textbf{k}\cdot\textbf{R}_\alpha-\phi_\textbf{k}) }\Delta_\alpha]$. 
 
The dispersion of the Bogoliubov quasiparticles can be solved exactly by diagonalizing Eq.~\eqref{eq:HamIntraInter} and is given by, 
\begin{equation}
 \begin{split}
E_{QP } &  = \pm \Bigg(\epsilon_\textbf{k}^2 +\mu^2 + |\Delta_\alpha|^2  \pm \sqrt{\Delta _\alpha^2 {\Delta _\alpha^\dagger}^2  \sin ^2(2 \textbf{k})  +4 \epsilon_\textbf{k}^2 (|\Delta _\alpha |^2 \sin ^2(\textbf{k}) +\mu ^2)}  \Bigg)^{1/2}
\end{split}
\end{equation}

As temperature approaches critical transition temperature $T_c$, the gap becomes small. Therefore, we can treat the gap perturbatively as
\begin{equation}
H_{MF}=H_0 +V = \begin{pmatrix}
\xi_1     &   0     &   0   &  0       \\
0      &   -\xi_1       & 0    &   0       \\
0  &  0   &   \xi_2     & 0 \\   
0      &   0       &0  &   -\xi_2     \\
\end{pmatrix} +
\begin{pmatrix}
0    &   \Delta_{i}     &   0   &   -\Delta_{I}       \\
\Delta_{i}^\dagger       &  0       & \Delta_{I}^\dagger     &   0       \\
0  &  \Delta_{I}   &   0    &   -\Delta_{i}  \\   
-\Delta_{I}^\dagger       &   0       &-\Delta_{i}^\dagger     &   0    \\
\end{pmatrix}
\end{equation}
Using the pertubation theory, one can write the new eigenstates ($|\Psi_{\alpha}\rangle$) in terms of eigenstates of $H_0$ ($|\Psi_{\alpha}^0 \rangle$) as
\begin{equation}
|\Psi_{\alpha}\rangle = |\Psi_{\alpha}^0\rangle + \sum_{\beta\neq \alpha}\frac{\langle \Psi_{\beta}^0| V_{\beta\alpha}   |\Psi_{\alpha}^0\rangle}{E_\alpha - E_\beta}  |\Psi_{\beta}^0\rangle
\end{equation}
The new eigenstates and eigenergies are as follows: 
\begin{equation}
\begin{split}
\gamma_1 & =  c_{k\uparrow}+\frac{\Delta_i}{2\xi_1} c_{-k\downarrow}^\dagger-\frac{\Delta_I}{\xi_1+\xi_2} d_{-k\downarrow}^\dagger,  ~~E_1 = \xi_1;  \qquad 
\gamma_2 = c_{-k\downarrow}^\dagger -\frac{\Delta_i^\dagger}{2\xi_1} c_{k\uparrow}-\frac{\Delta_I^\dagger}{\xi_1+ \xi_2} d_{k\uparrow},  ~~E_2 = -\xi_1; \\
\gamma_3 & =  d_{k\uparrow}-\frac{\Delta_i}{2\xi_2} d_{-k\downarrow}^\dagger + \frac{\Delta_I}{\xi_1+\xi_2} c_{-k\downarrow}^\dagger, ~~E_2 = \xi_2; \qquad 
\gamma_4 = d_{-k\downarrow}^\dagger +\frac{\Delta_i^\dagger}{2\xi_2} d_{k\uparrow} + \frac{\Delta_I^\dagger}{\xi_1+ \xi_2} c_{k\uparrow}, ~~E_4 = -\xi_2. \\
\end{split}
\end{equation}
Using the relation, that different component of  $\gamma_m$ are orthogonal, one can simplify the below equations,
 \begin{equation}
\begin{split}
\langle c_{k\uparrow} c_{-k\downarrow}\rangle  & = \frac{\Delta_i}{2\xi_1} (\gamma_1 \gamma_1^\dagger - \gamma_2  \gamma_2^\dagger) , \qquad
\langle d_{k\uparrow} d_{-k\downarrow}\rangle = \frac{\Delta_i}{2\xi_2}(-\gamma_3 \gamma_3^\dagger +\gamma_4  \gamma_4^\dagger), ~\\
\langle c_{k\uparrow} d_{-k\downarrow}\rangle  & = \frac{\Delta_I}{\xi_1+\xi_2} (-\gamma_1 \gamma_1^\dagger + \gamma_4 \gamma_4^\dagger) ,\qquad
 \langle d_{k\uparrow} c_{-k\downarrow}\rangle  = \frac{\Delta_I}{\xi_1+\xi_2}(\gamma_3 \gamma_3^\dagger -\gamma_2  \gamma_2^\dagger) 
\end{split}
\end{equation}
Bogoliubons follow the Fermi-Dirac distribution and their expectation value is given by $ \langle \gamma_m^\dagger \gamma_m\rangle = \frac{1}{1+e^{\beta E_m}} $. Therefore, one can write the pairing terms as,
 \begin{equation}
 \begin{split}
\langle c_{k\uparrow} c_{-k\downarrow}-  d_{k\uparrow}&  d_{-k\downarrow}\rangle  = \Delta_i \Big(\frac{\tanh(\tfrac{\beta\xi_1}{2}) }{2\xi_1}+ \frac{\tanh(\tfrac{\beta\xi_2}{2})}{2\xi_2}\Big),~
\langle d_{k\uparrow} c_{-k\downarrow} -  c_{k\uparrow}  d_{-k\downarrow}\rangle  = \frac{\Delta_I}{\xi_1+\xi_2} \Big(\tanh(\tfrac{\beta\xi_1}{2})+\tanh(\tfrac{\beta\xi_2}{2})\Big)
\end{split}
\end{equation}

Using the definition of the $\Delta_i$ and $\Delta_I$, the gap equation is simplified as,

\begin{equation}\label{eq:lnrzdGapEq}
\begin{split}
\Delta_\alpha & = J_\alpha \sum_k \cos(\textbf{k}\cdot\textbf{l}_\alpha-\phi_\textbf{k}) \langle c_{k\uparrow} c_{-k\downarrow}- \langle d_{k\uparrow} d_{-k\downarrow}\rangle - i\sin(\textbf{k}\cdot\textbf{l}_\alpha-\phi_\textbf{k}) \langle d_{k\uparrow} c_{-k\downarrow} - \langle c_{k\uparrow}d_{-k\downarrow}\rangle \\
 =  &  J_\alpha \sum_{k,\beta }\bigg[\cos(\textbf{k}\cdot\textbf{l}_\alpha-\phi_\textbf{k}) \cos(\textbf{k}\cdot \textbf{l}_\beta- \phi_\textbf{k})\Big(\frac{\tanh(\tfrac{\beta\xi_1}{2}) }{2\xi_1}+ \frac{\tanh(\tfrac{\beta\xi_2}{2})}{2\xi_2}\Big) \\
&+ \sin(\textbf{k}\cdot\textbf{l}_\alpha-\phi_\textbf{k}) \sin(\textbf{k}\cdot\textbf{l}_\beta- \phi_\textbf{k}) \frac{\big(\tanh(\tfrac{\beta\xi_1}{2})+\tanh(\tfrac{\beta\xi_2}{2}) \big)}{\xi_1+\xi_2}\bigg] \Delta_{\beta}
\end{split}
\end{equation}
The above linearized can be solved to evaluate the $T_c$ for the superconducting phases in the lattice. 

We are interested in the lattice with $D_\text{6h}$ and $D_{2h}$ symmetry with isotopic interaction along the three bonds. Therefore, we present results relevant to linearly polarized light polarized along $y$-direction discussed in the main text. In which the Hamiltonian has a reduced $D_\text{2h}$ symmetry with anisotorpic hoppings.  In the linearized gap equation this leads to anisotropy in the $\phi_{\textbf{k}}~(=\text{arg}(\sum_{\alpha} e^{i\textbf{k}\cdot\textbf{R}_\alpha} t_\alpha))$ and is reduced it to $D_\text{2h}$ symmetry due to reduced symmetry in the Hamiltonian. The linearized gap equation can be written in a  matrix form
\begin{equation}\label{eq:GapSelfConsistent}
\begin{bmatrix}
\Delta_{0}     \\
\Delta_{1}\\
\Delta_{2}   \\
\end{bmatrix}
=J \begin{pmatrix}
A_1		& 	B_1		&  B_1	\\   
B_1		& 	A_2		&  B_2	\\   
B_1		& 	B_2		&  A_2	\\   
\end{pmatrix}
\begin{bmatrix}
\Delta_{0}     \\
\Delta_{1}\\
\Delta_{2}   \\
\end{bmatrix}
\end{equation}
Here, $A_{\{1,2\}}$ and  $B_{\{1,2\}}$ are the corresponding prefactors of $\Delta_\beta$ in the Eq.~\eqref{eq:lnrzdGapEq}.  The solutions of the above equation is given by, $A_2 - B_2=\frac{1}{J}$ with the eigenvector: $(0,1,-1)/\sqrt{2}$) which is $d_{xy}$-wave. Another set of solutions are $\frac{1}{2}(A_1+A_2+B_2\pm \sqrt{(A_1-A_2-B_2)^2+8B_1^2} )=\frac{1}{J} $.  The eigenvectors of these solutions are a superposition of the $s$ and $d_{x^2-y^2}$-wave: $a_1 (2,-1,-1)/\sqrt{6}+a_2 (1,1,1)/\sqrt{2}$. The prefactor $a_1$ and $a_2$ are a function of $A_1, A_2, B_1$ and $B_2$. These results simplify to $s$, $d_{x^2-y^2}$ and $d_{xy}$-wave solution for the isotropic case preserving $D_\text{6h}$ symmetry as reported in Ref.~\cite{Black_Schaffer_2014}.

\end{appendix}

\twocolumngrid

 \bibliography{SWTSCHoneycomb}


\makeatletter
 	\title{Supplemental Material: ``Inducing and controlling superconductivity in Hubbard honeycomb model using an electromagnetic drive"}
 	
 	\author{Umesh Kumar}
 	\affiliation{Theoretical Division, Los Alamos National Laboratory, Los Alamos, New Mexico 87545, USA}
 	\author{Shi-Zeng Lin}
 	\affiliation{Theoretical Division, T-4 and CNLS, Los Alamos National Laboratory, Los Alamos, New Mexico 87545, USA}

 	\maketitle
 	In this supplemental, we present the mapping of time-dependent Hubbard model to the Floquet $t$-$J$ model. We start with a time-dependent Hamiltonian for a Hubbard honeycomb lattice given by 
 	\begin{equation}
 		\begin{split}
 			H(t) & =  \sum_{\langle ij \rangle,\sigma} t_{ij} e^{i\delta F_{ij}(t)}  c_{i\sigma}^\dagger c_{j\sigma}+  \text{h.c.}  + U \sum_{i} n_{i\uparrow} n_{i\downarrow}.
 		\end{split}
 	\end{equation}
 	Here, $t_{ij}$ is the hopping between nearest neighbour sites-$i, j$ and $U$ is the onsite electron repulsions. $\delta F_{ij}(t)$ is the Peierls phase given by $\delta F_{ij}(t) = \textbf{F}(t) \cdot (\textbf{r}_i-\textbf{r}_j)$ where $\textbf{F}(t)$ is the vector potential of the light. Also, the Hamiltonian is periodic in time with the interval $T$, $H(t+T) = H(t)$.
 	
 	\section{Schrieffer-Wolff transformation } \label{SWT}
 	We change to a rotating frame given by the transformation $\mathcal{H}(t) =  V^\dagger H(t) V -i V^\dagger \dot{V}$  using the unitary operator, $V(t) = e^{-i Ut\sum_{i} n_{i,\uparrow} n_{i,\downarrow} }$~\cite{PhysRevLett.116.125301, Eckardt_2015}. Here, $\hbar=1$.\\
 	
 	\\
 	The Hamiltonian in new gauge is given by
 	\begin{equation} \label{eq:TDHamiltonina}
 		\begin{split}
 			\mathcal{H}(t) & = e^{i Ut\sum_{i} n_{i\uparrow} n_{i\downarrow} } \sum_{\langle ij\rangle,\sigma} t_{ij} \big(e^{i\delta F_{ij}(t)} c_{i\sigma}^\dagger c_{j\sigma}+ \text{h.c.}  \big)e^{-i Ut\sum_{i} n_{i\uparrow} n_{i\downarrow} } \\ 
 			& = \sum_{\langle ij\rangle,\sigma} t_{ij}  [h_{i\bar{\sigma}} + e^{iUt} n_{i\bar{\sigma}}] \big( e^{i\delta F_{ij}(t)} c_{i\sigma}^\dagger c_{j\sigma} \big) [h_{j\bar{\sigma}} + e^{-iUt} n_{j\bar{\sigma}}] + \text{h.c.}
 		\end{split}
 	\end{equation}
 	In the above equation, we have used the relations; $c_{i,\sigma} n_{i,\sigma} = c_{i,\sigma},~ n_{i,\sigma} c_{i,\sigma}^\dagger  = c_{i,\sigma}^\dagger, ~  n_{i,\sigma} c_{i,\sigma} = 0,~ c_{i,\sigma}^\dagger n_{i,\sigma} =0$ and $h_{i\sigma} = 1-n_{i\sigma}$. One can further reorganize the above Hamiltonian ($i>j$) as
 	
 	\begin{equation}\label{eq:PierlswithU}
 		\begin{split}
 			\mathcal{H}(t) = \sum_{\langle ij\rangle,\sigma} & t_{ij} e^{i\delta F_{ij}(t)} g_{ij\sigma}+ \text{h.c.} + \sum_{\langle ij\rangle,\sigma} t_{ij}e^{i\delta F_{ij}(t) +iUt} h_{ij\sigma}^\dagger+t_{ij}e^{i\delta F_{ij}(t) -iUt} h_{ij\sigma}+\text{h.c.}
 		\end{split}
 	\end{equation}
 	The terms $g_{ij\sigma}$ and $h_{ij\sigma}$ in the above equation arise due to the restrictions imposed by the Hubbard term and are given by
 	\begin{equation}
 		\begin{split}
 			g_{ij\sigma} &  = h_{i\bar{\sigma}}c_{i\sigma}^\dagger c_{j,\sigma} h_{j\bar{\sigma}} + n_{i,\bar{\sigma}}c_{i,\sigma}^\dagger c_{j,\sigma}  n_{j,\bar{\sigma}} \\ 
 			h_{ij\sigma}  &  =  h_{i,\bar{\sigma}}c_{i\sigma}^\dagger c_{j\sigma}  n_{j\bar{\sigma}},  \qquad h_{ji\sigma}   =  h_{j,\bar{\sigma}}c_{j\sigma}^\dagger c_{i\sigma}  n_{i\bar{\sigma}}, 
 			\\ 
 			h_{ij\sigma}^\dagger  &  =  n_{i,\bar{\sigma}}c_{i\sigma}^\dagger c_{j\sigma}  h_{j\bar{\sigma}},  \qquad h_{ji\sigma}^\dagger   =  n_{j,\bar{\sigma}}c_{j\sigma}^\dagger c_{i\sigma}  h_{i\bar{\sigma}}  
 		\end{split}
 	\end{equation}
 	Notice that $g_{ij\sigma}^\dagger = g_{ji\sigma}$, but $h_{ij\sigma}^\dagger \neq h_{ji\sigma}$.
 	The first term in $g_{ij\sigma}$ denotes the hopping of holons, whereas the second term  denotes the doublons. $h_{ij\sigma}^\dagger$ denote hoping from a holon to doublon.  After deriving the Hamiltonian in the new gauge, one can map these Hamiltonian to a time-independent Floquet Hamiltonian. We consider the case of both cirularly and linearly polarized light to  map these onto Floquet $t$-$J$ model.

 	\subsection{Circularly Polarized Light}
 	We consider a circularly polarized light (CPL) for which vector potential is given by $\textbf{F}(t) =  \zeta[\sin(\omega t) \hat{e}_x - \cos(\omega t) \hat{e}_y]$, where $\zeta =A/\omega$. Peierls phase in the Hamiltonian, $\delta F_{ij}(t) = \textbf{F}(t) \cdot (\textbf{r}_i - \textbf{r}_j) =  \zeta\sin(\omega t- \phi_{ij})$, where $ \textbf{r}_i -\textbf{r}_j = \cos(\phi_{ij}) \hat{e}_x + \sin(\phi_{ij}) \hat{e}_y $.

 	To map the time-dependent Hamiltonian to the time-independent form, we make use of the time-periodicity and use the Fourier transform (FT) given by $H_l =\frac{1}{T} \int_{0}^{T}dt \mathcal{H}(t)e^{-i l \omega t}$. We Fourier transform each term in the Eq.~\eqref{eq:TDHamiltonina}. FT  of $g_{ij\sigma}$ is given by
 	\begin{equation}
 		\begin{split}
 			\frac{1}{T} \int_{0}^{T}dt e^{i\delta F_{ij}(t)} e^{-i l \omega t} g_{ij\sigma}  & =  \frac{1}{T} \int_{0}^{T}dt e^{i(\zeta\sin(\omega t- \phi_{ij})- l \omega t)} g_{ij\sigma}  = e^{-il\phi_{ij}}\mathcal{J}_l g_{ij\sigma} 
 		\end{split}
 	\end{equation}
 	For the FT of $h_{ij\sigma}$, one consider $U=k\omega$, with $k$ and $l$ being co-prime. FT is then given by
 	\begin{equation}
 		\begin{split}
 			\frac{1}{T} \int_{0}^{T}dt e^{i\delta F_{ij}(t)+iUt} e^{-i l \omega t} h_{ij\sigma}^\dagger  & =  \frac{1}{T} \int_{0}^{T}dt e^{i(\zeta\sin(\omega t- \phi_{ij})- (l-k) \omega t)} h_{ij\sigma}^\dagger = e^{-i(l-k)\phi_{ij}}\mathcal{J}_{l-k} h_{ij\sigma}^\dagger
 		\end{split}
 	\end{equation}
 	
 	In the above equations, $\mathcal{J}_l(\zeta)= {\frac {1}{T}}\int _{0 }^{T}e^{i(\zeta \sin (\omega t) -l\omega t )}\,d \omega t $ is the Bessel functions of first kind and hence follows the relation $\mathcal{J}_{-l} (\zeta) = (-1)^l \mathcal{J}_{l}(\zeta)$. Hamiltonian in Eq.~\eqref{eq:PierlswithU} can hence be written in the Fourier basis given by
 	\begin{equation}\label{Eq:Hl}
 		\begin{split}
 			H_l = \sum_{\langle ij\rangle,\sigma} & \mathcal{J}_l (\zeta)  e^{-il\phi_{ij}} g_{ij\sigma} +\mathcal{J}_{-l}(\zeta)   e^{-il\phi_{ij}} g_{ij\sigma}^\dagger
 			+ \mathcal{J}_{l-k}(\zeta)   e^{-i(l-k)\phi_{ij}} h_{ij\sigma}^\dagger + \mathcal{J}_{l+k} (\zeta)  e^{-i(l+k)\phi_{ij}} h_{ij\sigma} \\
 			+ & \mathcal{J}_{-l+k} (\zeta)  e^{-i(l-k)\phi_{ij}} h_{ji\sigma}^\dagger    + \mathcal{J}_{-l-k}(\zeta)   e^{-i(l+k)\phi_{ij}} h_{ji\sigma}
 		\end{split}
 	\end{equation}
 	For the sake of convenience, we have have dropped the prefactors ($t_{ij}$), but will retrieve it in final form of the Hamiltonian.  \\
 	
 	{\bf High-frequency expansion:---} In the high-frequency limit, the effective time-independent Floquet Hamiltonian~\cite{PhysRevLett.116.125301} can be written as,
 	\begin{equation}
 		\begin{split}
 			H_\text{eff} & = \sum_{n=0}^{\infty} H_\text{eff}^{(n)},\quad ~~ 
 			\text{where},  \quad H_\text{eff}^{(0)}   = H_{l=0},  \qquad
 			H_\text{eff}^{(1)}   =  \sum_{l=1}^{\infty} \frac{1}{l\omega} [H_l, H_{-l}]
 		\end{split}
 	\end{equation} 
 	The high order ($N$) terms are of the order ($1/\omega^N$) and hence allows for simplification. In our our work here, we evaluate only the leading correction.
 	Also, we are interested in the low energy dynamics, we therefore, use a projector $P_0=\Pi_{i=1}^N (1-n_{i,\uparrow} n_{i,\downarrow})$ to remove doubly occupied states. The zeroth order term is given by
 	\begin{equation}
 		\begin{split}
 			P_0 H_\text{eff}^{(0)}  P_0 = P_0 H_{l=0}  P_0=\mathcal{J}_0 \sum_{\langle ij \rangle,\sigma}  (h_{i,\bar{\sigma}})c_{i,\sigma}^\dagger c_{j,\sigma} (h_{j,\bar{\sigma}}) 
 		\end{split}
 	\end{equation} 
 	To evaluate the leading correction, we need to evaluate $P_0 [H_, H_{-l}]P_0/l$. The term corresponding to $g_{ij\sigma}$ simplifies as, 
 	\begin{equation}
 		\begin{split}
 			& P_0\sum_{\substack{\langle ij \rangle ,\sigma \\ \langle i'j' \rangle,\sigma'}}[\mathcal{J}_l(\zeta) e^{-il\phi_{i'j'}}g_{i'j'\sigma'}, \mathcal{J}_{-l}(\zeta) e^{il\phi_{ij}} g_{ij\sigma}] P_0= \mathcal{J}_{l}(\zeta)  \mathcal{J}_{-l}(\zeta)  \sum_{\langle i'ij\rangle ,\sigma} [-2i\sin(l[\phi_{i'i}-\phi_{ij}])]  h_{i',\bar{\sigma}}c_{i',\sigma}^\dagger h_{i,\bar{\sigma}} c_{j,\sigma} h_{j,\bar{\sigma}} 
 		\end{split}
 	\end{equation} 
 	Here $i'$ and $j$ are next nearest neighbours (NNN) and similarly the other terms corresponding to $g_{ij\sigma}$ can be evaluated. 
 	
 	In the term corresponding to doublon to holon and vice versa hopping,  only the terms without any double occupancy survive. We here present terms of the form, $h_{ji\sigma}h_{ij\sigma'}^\dagger$ and $h_{i'i\sigma}h_{ij\sigma'}^\dagger$. There are two possibilities:  a) $i'=i$ and $j'=j$  (two-sites)
 	\begin{equation}\label{eq:hijhji}
 		\begin{split}
 			&   P_0\sum_{\substack{\langle ij\rangle,\sigma, \\  \langle i'j'\rangle,\sigma'}}[\mathcal{J}_{l+k}(\zeta)e^{-i(l+k)\phi_{i'j'}}h_{j'i'\sigma'}, \mathcal{J}_{-l-k}(\zeta) e^{-i(-l-k)\phi_{ij}} h_{ij\sigma}^\dagger] P_0=  + (-1)^{l+k} \mathcal{J}_{l+k}^2(\zeta)\sum_{\langle ij\rangle,\sigma,\sigma'} h_{ji\sigma'} h_{ij\sigma}^\dagger 
 			\\&  
 			P_0\sum_{\substack{\langle ij\rangle,\sigma, \\  \langle ij\rangle,\sigma'}}[ \mathcal{J}_{l-k}(\zeta) e^{-i(l-k)\phi_{ij}} h_{ij\sigma}^\dagger, \mathcal{J}_{-l+k}(\zeta) e^{-i(-l+k)\phi_{i'j'}}h_{j'i'\sigma'}] P_0= - (-1)^{l-k}\mathcal{J}_{l-k}^2(\zeta) \sum_{\langle ij\rangle,\sigma,\sigma'}  h_{ji\sigma'} h_{ij\sigma}^\dagger 
 		\end{split}
 	\end{equation} 
 	and b) $j'=i$ (three-sites)
 	\begin{equation}
 		\begin{split}
 			&   P_0\sum_{\substack{\langle ij\rangle,\sigma, \\  \langle i'j'\rangle,\sigma'}}  [\mathcal{J}_{l+k}(\zeta) e^{-i(l+k)\phi_{i'j'}} h_{i'j'\sigma'}, \mathcal{J}_{-l-k}(\zeta) e^{-i(-l-k)\phi_{ij}} h_{ij\sigma}^\dagger] P_0 = +  (-1)^{l+k} \mathcal{J}_{l+k}^2 \sum_{\langle i'ij\rangle,\sigma,\sigma'} e^{-i(l+k)(\phi_{i'i}- \phi_{ij})} h_{i'i\sigma'} h_{ij\sigma}^\dagger
 			\\
 			&  P_0\sum_{\substack{\langle ij\rangle,\sigma, \\  \langle i'j'\rangle,\sigma'}}  [\mathcal{J}_{l-k}(\zeta) e^{-i(l-k)\phi_{ij}} h_{ij\sigma}^\dagger, \mathcal{J}_{-l+k}(\zeta) e^{-i(-l+k)\phi_{i'j'}}h_{i'j'\sigma}] P_0 = -(-1)^{l-k} \mathcal{J}_{l-k}^2 \sum_{\langle i'ij\rangle,\sigma,\sigma'} e^{i(l-k)(\phi_{i'i}-\phi_{ij})} h_{i'i\sigma'} h_{ij\sigma}^\dagger 
 		\end{split}
 	\end{equation} 
 	Similarly, one can evaluate the expression of the form, $h_{ij\sigma} h_{ji\sigma'}^\dagger$ and $h_{i'j\sigma}h_{ji\sigma'}^\dagger$. 
 	
 	Using the above expression for the term in the Fourier basis, the first-order correction is given by 
 	\begin{equation}
 		\begin{split}
 			P_0\frac{[H_l, H_{-l}]}{l\omega } P_0 & = -(-1)^l(\mathcal{J}_l)^2 \Big(\frac{2i\sin(l[\phi_{i'i}-\phi_{ij}])}{l\omega}h_{i',\bar{\sigma}} c_{i',\sigma}^\dagger h_{i,\bar{\sigma}}c_{j,\sigma}h_{j,\bar{\sigma}}+\frac{2i\sin(l[\phi_{j'j}-\phi_{ij}])}{l\omega}h_{j',\bar{\sigma}} c_{j',\sigma}^\dagger h_{j,\bar{\sigma}} c_{i,\sigma}h_{i,\bar{\sigma}}\Big)  \\
 			&   + \frac{(\mathcal{J}_{l+k})^2}{l\omega}  h_{ji\sigma'} h_{ij\sigma}^\dagger -\underbrace{ \frac{(\mathcal{J}_{l-k})^2}{l\omega} }_{l\rightarrow -l}  h_{ji\sigma'} h_{ij\sigma}^\dagger + \frac{(\mathcal{J}_{l+k})^2}{l\omega}  h_{ij\sigma'} h_{ji\sigma}^\dagger - \underbrace{\frac{(\mathcal{J}_{-l+k})^2}{l\omega} }_{l\rightarrow -l} h_{ij\sigma'} h_{ji\sigma}^\dagger  ~ [\text{2-sites}]\\ 
 			&  + ( (-1)^{l+k} \frac{\mathcal{J}_{l+k}^2 }{l\omega} e^{-i(l+k)(\phi_{i'i}-\phi_{ij})}-\underbrace{  (-1)^{l-k} \frac{\mathcal{J}_{l-k}^2  }{l\omega}  e^{-i(l-k)(\phi_{ij}-\phi_{i'i})} }_{l\rightarrow -l}) h_{i'i\sigma'} h_{ij\sigma}^\dagger  \quad [\text{3-sites}]\\
 			&  + (-1)^{l+k} \frac{\mathcal{J}_{l+k}^2  }{l\omega} e^{-i(l+k)(\phi_{jj'}- \phi_{ij})} -\underbrace{ (-1)^{l-k} \frac{\mathcal{J}_{l-k}^2 }{l\omega} e^{-i(l-k)(\phi_{ij}- \phi_{jj'})}}_{l\rightarrow -l})  h_{j'j\sigma'} h_{ji\sigma}^\dagger\\
 		\end{split}
 	\end{equation} 
 	Notice that in the above equation, the substitution $l\rightarrow -l$, allows one to change the summation over $l$ from $l\in [1,\infty)$ to $l \in \{(-\infty, \infty) -0\}$ and also remember $\mathcal{J}_{-l}(\zeta) = (-1)^{l} \mathcal{J}_{l}(\zeta)$. One can therefore, rewrite the first order correction as
 	\begin{equation}
 		\begin{split}
 			\sum_{l=1}^{\infty}P_0\frac{[H_l, H_{-l}]}{l\omega } P_0  
 			&= \sum_{\langle ij\rangle, \sigma, \sigma'}  \sum_{\substack{l=-\infty,\\ l\neq 0}}^{\infty}  - (\mathcal{J}_l)^2  \frac{i\sin(l[\phi_{i'i}-\phi_{ij}+\pi])}{l\omega}h_{i,\bar{\sigma}} c_{i,\sigma}^\dagger h_{j,\bar{\sigma}} c_{j',\sigma}h_{j',\bar{\sigma}} + h.c. + 
 			\frac{(\mathcal{J}_{l+k})^2}{l\omega}  h_{ij\sigma} h_{ji\sigma}^\dagger
 			\\
 			&+
 			\frac{(\mathcal{J}_{l+k})^2}{l\omega}  h_{ji\sigma} h_{ij\sigma}^\dagger  
 			+ \frac{(\mathcal{J}_{l+k})^2}{l\omega}(e^{-i(l+k)(\phi_{i'i}- \phi_{ij}-\pi)} h_{i'i\sigma'} h_{ij\sigma}^\dagger + e^{-i(l+k)(\phi_{jj'}- \phi_{ij}+\pi)} h_{j'j\sigma'} h_{ji\sigma}^\dagger)  \\
 		\end{split}
 	\end{equation} 
 	
 	In the honeycomb lattice, $\phi_{i'i} -\phi_{ij} = +(-)~\pi/3$ and $\phi_{jj'} -\phi_{ij} = +(-)~\pi/3$ for the anticlockwise(clockwise) next-nearest hopping given by $\tau_{ij} = +(-)$.  Also, using the substitution $l+k \rightarrow -l'$ and $-l' \rightarrow l'$. We can then rewrite the above equation as
 	\begin{equation}\label{eq:hc_Heff1}
 		\begin{split}
 			\sum_{l=1}^{\infty}P_0\frac{[H_l, H_{-l}]}{l\omega } P_0 &  = \sum_{\langle ij\rangle, \sigma, \sigma'} \Bigg[ \sum_{\substack{l=-\infty,\\ l\neq 0}}^{\infty} \tau_{ij} (\mathcal{J}_l)^2   \frac{i\sin(2\pi l/3)}{l\omega}h_{i,\bar{\sigma}} c_{i,\sigma}^\dagger c_{j,\sigma}h_{j,\bar{\sigma}}+h.c. - \sum_{\substack{l'=-\infty, \\ l' \neq -k}}^{\infty}  \frac{(\mathcal{J}_{l'})^2}{l'\omega+U} (h_{ij\sigma} h_{ji\sigma}^\dagger+ h_{ji\sigma} h_{ij\sigma}^\dagger ) \\
 			& - \sum_{\substack{l'=-\infty, \\ l'\neq -k}}^{\infty} \tau_{ij} \frac{(\mathcal{J}_{l'} )^2}{l'\omega+U}[\cos(2\pi l'/3) (h_{i'i\sigma'} h_{ij\sigma}^\dagger + h_{j'j\sigma'} h_{ji\sigma}^\dagger) +i\sin(2\pi l'/3) (h_{ij\sigma} h_{ji\sigma}^\dagger+ h_{ji\sigma} h_{ij\sigma}^\dagger ) ]\Bigg]
 		\end{split}
 	\end{equation} 
 	
 	We make use the relations given below to rewrite the Hamiltonian into a $t$-$J$ model~\cite{PhysRevB.37.9753,SpalektJ}
 	\begin{equation}\label{eq:t1t2Jstatic}
 		\begin{split}
 			\sum_{\langle ij\rangle, \sigma, \sigma'}(h_{ij\sigma'} h_{ji\sigma}^\dagger+ h_{ji\sigma'} h_{ij\sigma}^\dagger )  & = -2\sum_{ij\sigma} 
 			(\textbf{S}_i \cdot \textbf{S}_j - \sum_{\sigma'} \frac{1}{4}h_{i,\bar{\sigma}}n_{i\sigma} n_{j\sigma'} h_{j\bar{\sigma}}) \quad \quad ~[2-\text{sites}]  \\ 
 			\sum_{\langle ij\rangle, \sigma, \sigma'}(h_{i'i\sigma'} h_{ij\sigma}^\dagger+ h_{j'j\sigma'} h_{ji\sigma}^\dagger ) & = - \sum_{ikj,\sigma} ~[h_{i\bar{\sigma}}c_{i,\sigma}^\dagger n_{k\bar{\sigma}}h_{k\sigma} c_{j\sigma}h_{j\bar{\sigma}}-h_{i\bar{\sigma}}c_{i,\sigma}^\dagger (c_{k\bar{\sigma}}^\dagger c_{k\sigma})h_{k\bar{\sigma}} c_{j\bar{\sigma}}h_{j\sigma}] \quad [3-\text{sites}]
 		\end{split}
 	\end{equation} 
 	In the above equation, the first term corresponds to the double spin flips and the second term corresponds to three-site hopping both with and without spin-flips. We notice that the leading corrections given by Eq.~\eqref{eq:hc_Heff1} are of the order of $1/\omega$. Further, the first terms of NNN hopping does not contain the term $l=0$, which is the only Bessel term that contributes for small $\zeta$, this can become important when $\mathcal{J}_0({\zeta})$ vanishes~\cite{PhysRevB.93.144307}. Also, the three-site hopping given by the is ignored for the sake of simplicity as is usually the case in these downfolded $t$-$J$ models at low doping. 
 	In this work, we limit ourselves in the small $\zeta$ regime. 
 	We therefore use the high-frequency approximation which leads to two set of cases: 
 	
 	a) $t\ll U\ll \omega$: In this case only the $l'=0$ contribute to the leading correction term in Eq.~\eqref{eq:hc_Heff1}, and the Floquet Hamiltonian is given by, 
 	\begin{equation} 
 		\begin{split}
 			H_\text{F}^{t\ll U \ll \omega}  & \approx  \mathcal{J}_0(\zeta ) \sum_{ij,\sigma} t_{ij} h_{i,\bar{\sigma}}c_{i,\sigma}^\dagger c_{j,\sigma} h_{j,\bar{\sigma}} + \sum_{ij, \sigma}  (\mathcal{J}_{0}(\zeta))^2 \frac{2t_{ij}^2}{U}  (\textbf{S}_i \cdot \textbf{S}_j -  \sum_{\sigma'} \frac{1}{4}h_{i,\bar{\sigma}}n_{i\sigma} n_{j\sigma'} h_{j\bar{\sigma}})
 		\end{split}
 	\end{equation} 
 	
 	b) $t\ll\omega\ll U$: In this case, since $U$ is the larger energy scale, we use the approximation $l\omega/U \ll 1$ and the relation $\sum_{l=-\infty}^{\infty}(\mathcal{J}_{l})^2  =1 $. The leading correction is given by 
 	\begin{equation} 
 		\begin{split}
 			H_\text{eff}^{(1)}   \approx & \sum_{ij, \sigma, \sigma'} -  \frac{2t_{ij}^2}{U} (h_{ij\sigma'} h_{ji\sigma}^\dagger+ h_{ji\sigma} h_{ij\sigma'}^\dagger )\sum_{\substack{l'=-\infty, \\ l' \neq -k}}^{\infty}  \frac{(\mathcal{J}_{l'})^2}{1+l'\omega/U} = \sum_{ij\sigma} \frac{2t_{ij}^2}{U}  
 			(\textbf{S}_i \cdot \textbf{S}_j - \sum_{\sigma'} \frac{1}{4}h_{i,\bar{\sigma}}n_{i\sigma} n_{j\sigma'} h_{j\bar{\sigma}}) 
 		\end{split}
 	\end{equation} 
 	Therefore, in this limit, the Floquet Hamiltonian is given by 
 	\begin{equation}\label{eq:CPL_FltJ} 
 		\begin{split}
 			H^F  & \approx  \mathcal{J}_0(\zeta ) \sum_{i,j,\sigma} t_{ij} h_{i,\bar{\sigma}}c_{i,\sigma}^\dagger c_{j,\sigma} h_{j,\bar{\sigma}} + \sum_{ij\sigma} \frac{2t_{0}^2}{U}  (\textbf{S}_i \cdot \textbf{S}_j - \sum_{\sigma'} \frac{1}{4}h_{i,\bar{\sigma}}n_{i\sigma} n_{j\sigma'} h_{j\bar{\sigma}})
 		\end{split}
 	\end{equation} 
 	In above equation, with the definition of superexchange given by $J = \frac{2t_{0}^2}{U} $, we have presented a simplified time-independent Floquet $t$-$J$ model derived from the time-dependent Hubbard model.

 	\subsection{Linerarly polarized light}
 	We consider a linearly polarized light along the $y$-direction for which the vector potential is given by   $\textbf{F}(t) =  \zeta\sin(\omega t) \hat{e}_y $, where $\zeta =A/\omega$. Peierls phase is given by $\delta F(t) =\textbf{F}(t)\cdot (\textbf{r}_i-\textbf{r}_j) = \zeta \sin(\omega t) \sin(\phi_{ij}) $ where $\textbf{r}_i-\textbf{r}_j = \cos({\phi_{ij}})\hat{e}_x+ \sin({\phi_{ij}})\hat{e}_y$. 
 	

 	To map the time-dependent Hamiltonian in Eq.~\eqref{eq:TDHamiltonina} onto a time-independent form,
 	we Fourier transform (FT) the rotated Hamiltonian given by $H_l =\frac{1}{T} \int_{0}^{T}dt \mathcal{H}(t)e^{-i l \omega t}$.  FT of the $g_{ij\sigma}$ term is given by
 	\begin{equation}
 		\begin{split}
 			\frac{1}{T} \int_{0}^{T}dt e^{i\delta F_{ij}(t)} e^{-i l \omega t} g_{ij\sigma}  & =  \frac{1}{T} \int_{0}^{T}dt e^{i(\zeta\sin(\omega t)\sin(\phi_{ij})- l \omega t)} g_{ij\sigma}  = \mathcal{J}_l(\zeta  \sin(\phi_{ij})) g_{i_\alpha j\sigma} 
 		\end{split}
 	\end{equation}
 	
 	
 	Whereas, for the FT of terms of form $h_{ij\sigma}$, one assumes $U=k\omega$, with $k$ and $l$ being co-prime. The FT in this case is given by
 	\begin{equation}
 		\begin{split}
 			\frac{1}{T} \int_{0}^{T}dt e^{i\delta F_{ij}(t)+iUt} e^{-i l \omega t} h_{ij\sigma}^\dagger  & =  \frac{1}{T} \int_{0}^{T}dt e^{i(\zeta\sin(\omega t)\sin(\phi_{ij})- (l-k) \omega t)} h_{ij\sigma}^\dagger = \mathcal{J}_{l-k}(\zeta\sin(\phi_{ij})) h_{i_\alpha j\sigma}^\dagger
 		\end{split}
 	\end{equation}
 	The three nearest neighbours for every site are given by $\phi_{i_\ell j} = \pi/2+2\pi \ell/3$ for $\ell =\{0,1,2\}$. The magnitude of the phase factors along the three bonds are anisotropic  and are given by $\sin(\phi_{i_\ell j}) = \{ 1, -\frac{1}{2}, -\frac{1}{2} \}$ for the three bonds $\ell =\{0,1,2\}$ respectively.
 	Finally, one can write the Fourier-transformed Hamiltonian as
 	\begin{equation}\label{Eq:Hl2}
 		\begin{split}
 			H_l & = \sum_{ij,\sigma} \sum_{\ell = 0,1,2}  \Big[ \mathcal{J}_l^\ell  g_{i_\ell j\sigma} +\mathcal{J}_{-l}^\ell  g_{i_\ell j\sigma}^\dagger
 			+ \mathcal{J}_{l-k}^\ell  h_{i_\ell j\sigma}^\dagger +  \mathcal{J}_{-l+k}^\ell   h_{ji_\ell\sigma}^\dagger   + \mathcal{J}_{l+k}^\ell   h_{i_\ell j\sigma} + \mathcal{J}_{-l-k}^\ell h_{ji_\ell\sigma} \Big]
 			\\
 			& =\sum_{ij,\sigma} \Big[ \mathcal{J}_l(\zeta)  g_{i_0 j\sigma} +\mathcal{J}_{-l}(\zeta)  g_{i_0 j\sigma}^\dagger
 			+ \mathcal{J}_{l-k}(\zeta)  h_{i_0 j\sigma}^\dagger +  \mathcal{J}_{-l+k}(\zeta)   h_{ji_0\sigma}^\dagger   + \mathcal{J}_{l+k}(\zeta)   h_{i_0 j\sigma} + \mathcal{J}_{-l-k}(\zeta) h_{ji_0\sigma} 
 			\\&  
 			+ \sum_{\ell=1,2}\mathcal{J}_{-l}(\zeta/2)  g_{i_\ell j\sigma} +\mathcal{J}_{l}(\zeta/2)  g_{i_\ell j\sigma}^\dagger
 			+ \mathcal{J}_{-l+k}(\zeta/2)  h_{i_\ell j\sigma}^\dagger +  \mathcal{J}_{l-k}(\zeta/2)   h_{ji_\ell\sigma}^\dagger   + \mathcal{J}_{-l-k}(\zeta/2)   h_{i_\ell j\sigma} + \mathcal{J}_{l+k}(\zeta/2) h_{ji_\ell\sigma}  \Big]
 		\end{split}
 	\end{equation}
 	and here $\mathcal{J}_l^\ell  = \mathcal{J}_l(\zeta\sin(\phi_{i_{\ell}j}))$. Following the discussion in CPL case, one can write the time-independent equation using high-frequency approximation.  Since, we are only interested in the singly occupied sector, we project the Hamiltonian with projector $P_0=\Pi_{i=1}^N (1-n_{i,\uparrow} n_{i,\downarrow})$ and the zeroth order term is given by
 	\begin{equation}\label{eq:LPLHeff_0}
 		\begin{split}
 			P_0 H_\text{eff}^{(0)}  P_0 = P_0 H_{l=0}  P_0=\sum_{ij,\sigma}  \big[\mathcal{J}_0(\zeta)  h_{i_0,\bar{\sigma}}c_{i_0,\sigma}^\dagger c_{j,\sigma} h_{j,\bar{\sigma}}+\sum_{\ell =1,2} \mathcal{J}_0(\zeta/2)  h_{i_\ell,\bar{\sigma}}c_{i_\ell,\sigma}^\dagger c_{j,\sigma} h_{j,\bar{\sigma}}	+ \text{h.c.} \big]
 		\end{split}
 	\end{equation} 
 	
 	The leading correction is given by $[H_l, H_{-l}]/l$. Since, there is no phase factor associated with the terms, the chiral NNN hopping term  vanishes. In this case also, we neglect the three site-hopping term as in the case of CPL. The commutation for superexchnage terms are given by
 	\begin{equation}\label{eq:hijhji}
 		\begin{split}
 			& P_0\sum_{\substack{ij\sigma, \\  i' j'\sigma'}}[\mathcal{J}_{-l-k}^\ell h_{j'i'_{\ell}\sigma'}, \mathcal{J}_{-l-k}^\ell h_{i_{\ell}j\sigma}^\dagger] P_0= \sum_{\substack{ij,\sigma,\sigma'}} +[\mathcal{J}_{-l-k}(\zeta\sin(\phi_{i_\ell j}))]^2 h_{ji\sigma'} h_{ij\sigma}^\dagger
 			\\
 			& P_0\sum_{\substack{ij,\sigma, \\  i'j',\sigma'}}[\mathcal{J}_{l-k}^\ell h_{i_\ell j\sigma}^\dagger, \mathcal{J}_{l-k}^\ell h_{j'i'_\ell \sigma'}] P_0= \sum_{\substack{ij,\sigma,\sigma'}} -[\mathcal{J}_{l-k}(\zeta\sin(\phi_{i_\ell j})) ]^2 h_{ji_\ell \sigma'} h_{i_\ell j\sigma}^\dagger \qquad [j'=j,~i'=i]
 		\end{split}
 	\end{equation} 
 	The leading order correction without the three site terms is given by 
 	\begin{equation}
 		\begin{split}
 			\sum_{l=1}^{\infty}P_0\frac{[H_l, H_{-l}]}{l\omega } P_0  &  =   \sum_{\substack{l=-\infty,\\ l'\neq -k}}^{\infty} \sum_{\substack{i j,\sigma,\sigma'}} 
 			-(\mathcal{J}_{l'}^\ell)^2 \frac{t_{ij}}{l'\omega+U}  (h_{i_{\ell}j\sigma'} h_{ji_{\ell}\sigma}^\dagger + h_{ji_{\ell}\sigma'} h_{i_{\ell}j\sigma}^\dagger) 
 		\end{split}
 	\end{equation} 
 	
 	Explicitly, one can write the leading correction term as 
 	\begin{equation} 
 		\begin{split}
 			H_\text{eff}^{(1)}   \approx \sum_{\substack{i j,\sigma,\sigma'}}   -  \frac{t_{ij}^2}{U} \Bigg[(& h_{i_0j\sigma} h_{ji_0\sigma}^\dagger+ h_{ji_0 \sigma} h_{i_0 j\sigma}^\dagger )\sum_{\substack{l'=-\infty, \\ l' \neq -k}}^{\infty}  \frac{(\mathcal{J}_{l'}(\zeta))^2}{1+l'\omega/U} +\sum_{\ell =1,2}(h_{i_\ell j\sigma} h_{ji_\ell\sigma}^\dagger+ h_{ji_\ell \sigma} h_{i_\ell j\sigma}^\dagger )\sum_{\substack{l'=-\infty, \\ l' \neq -k}}^{\infty}  \frac{(\mathcal{J}_{l'}(\zeta/2))^2}{1+l'\omega/U}  \Bigg]
 		\end{split}
 	\end{equation} 
 	We limit ourselves in the small $\zeta$ regime as was the case in CPL. In the high-frequency approximation depending on the $U$ and $\omega$, we have 
 	two cases:
 	
 	a) $t\ll U\ll \omega$: in this case, only the $l'=0$ term contribute to the terms with $U$ and the leading correction is given by 
 	\begin{equation} 
 		\begin{split}
 			H_\text{eff}^{(1)}   \approx  \sum_{\substack{i j,\sigma,\sigma'}}  -  &\frac{2t_{ij}^2}{U} \Big[(\mathcal{J}_{0}(\zeta))^2 ( h_{i_0j\sigma} h_{ji_0\sigma}^\dagger+ h_{ji_0 \sigma} h_{i_0 j\sigma}^\dagger )+ \sum_{\ell =1,2}(\mathcal{J}_{0}(\zeta/2))^2 (h_{i_\ell j\sigma} h_{ji_\ell \sigma}^\dagger+ h_{ji_\ell \sigma} h_{i_\ell j\sigma}^\dagger ) \Big].
 		\end{split}
 	\end{equation} 
 	
 	Using Eq.~\eqref{eq:t1t2Jstatic}, one can write the above equations in as a spin Hamiltonian. The leading correction is given by
 	\begin{equation} 
 		\begin{split}
 			H_\text{eff}^{(1)}  & \approx \sum_{\substack{i j,\sigma}} \frac{2t_{ij}^2}{U} \Big[(\mathcal{J}_{0}(\zeta))^2 (\textbf{S}_{i_0} \cdot \textbf{S}_j -  \sum_{\sigma'} \frac{1}{4}h_{i_0,\bar{\sigma}}n_{i_0\sigma} n_{j\sigma'} h_{j,\bar{\sigma'}})+ \sum_{\ell =1,2} (\mathcal{J}_{0}(\zeta/2))^2 (\textbf{S}_{i_\ell} \cdot \textbf{S}_j -   \sum_{\sigma'} \frac{1}{4}h_{i_\ell,\bar{\sigma}}n_{i_\ell\sigma} n_{j\sigma'} h_{j,\bar{\sigma}'})) \Big]
 		\end{split}
 	\end{equation}

 	b) $t\ll \omega\ll U$: In this case, we use the approximation $l\omega/U \ll 1$ and the identity $\sum_{l=-\infty}^{\infty} \mathcal{J}_l(\zeta)\mathcal{J}_l(\zeta)=1$ for all $\zeta$, and $\sum_{l=-\infty}^{\infty} \mathcal{J}_l(\zeta)\mathcal{J}_l(\zeta/2)\rightarrow 1$ for small $\zeta$. Here, the leading correction is given by
 	\begin{equation} 
 		\begin{split}
 			H_\text{eff}^{(1)} \approx \sum_{\substack{i j,\sigma}} \frac{2t_{ij}^2}{U} \Big[\sum_{\ell =0, 1,2} (\textbf{S}_{i_\ell} \cdot \textbf{S}_j -   \sum_{\sigma'} \frac{1}{4}h_{i_\ell,\bar{\sigma}}n_{i_\ell\sigma} n_{j\sigma'} h_{j,\bar{\sigma}'} \Big].
 		\end{split}
 	\end{equation} 
 	
 	Therefore, using the above equations and Eq.~\eqref{eq:LPLHeff_0}, the $t$-$J$ Floquet Hamiltonian is given by  
 	
 	\begin{equation}
 		\begin{split}
 			H_{F} =\sum_{ij,\sigma}  t_{ij} \big[\mathcal{J}_0(\zeta)  h_{i_0,\bar{\sigma}}c_{i_0,\sigma}^\dagger c_{j,\sigma} h_{j,\bar{\sigma}} +\sum_{\ell =1,2} \mathcal{J}_0(\zeta/2)  h_{i_\ell,\bar{\sigma}}c_{i_\ell,\sigma}^\dagger c_{j,\sigma} h_{j,\bar{\sigma}}	+ \text{h.c.}  \big]+H_\text{eff}^{(1)}
 		\end{split}
 	\end{equation}
 	From the above equations one can see that that in the limit $t\ll U\ll \omega$, superexchange is modified whereas, in the limit $t\ll \omega\ll U$, it is unaffected. On the other hand, hopping is rescaled by the Bessel function in both the cases.


\makeatother
\myexternaldocument{Supplement}

\end{document}